\documentclass[a4paper]{article}
\usepackage{etoolbox}
\newbool{PREPRINT}
\booltrue{PREPRINT}

\ifbool{PREPRINT}{ 
\usepackage[colorlinks,bookmarksopen,bookmarksnumbered,citecolor=red,urlcolor=red]{hyperref}
\usepackage[affil-it]{authblk}
\usepackage[cm]{fullpage}
\usepackage{amsthm}

}{} 
\usepackage{amsmath,amssymb}
\usepackage{fullpage}
\usepackage{algorithm,algcompatible}
\renewcommand{\COMMENT}[2][.7\linewidth]{%
  \leavevmode\hfill\makebox[#1][l]{#2}}
\usepackage{listings}
\usepackage{color}
\usepackage{graphicx}
\usepackage{multirow}
\usepackage{subfig}
\usepackage{mathrsfs}

\newcommand{\figdir}{./}
\renewcommand{\vec}[1]{\ensuremath{\boldsymbol{#1}}}
\newcommand{\zhat}{\hat{\vec{z}}}
\newcommand{\ddt}[1]{\frac{\partial #1}{\partial t}}

\newcommand{\order}{\mathcal{O}}
\newcommand{\Uspace}{\mathbb{U}}
\newcommand{\Vspace}{\mathbb{V}}
\newcommand{\Wspace}{\mathbb{W}}
\newcommand{\Hdiv}{\texttt{HDiv}}

\newcommand{\nuCFL}{\ensuremath{\nu_{\operatorname{CFL}}}}

\lstdefinelanguage[firedrake]{python}[]{python}{%
  emph={[2]Function,MeshHierarchy,FunctionSpaceHierarchy,restrict,prolong,
    assemble,solve,Kernel,par_loop},
  emph={grad,dx,dot,action,op2}
}
\definecolor{DarkBlue}{rgb}{0.00,0.00,0.55}
\definecolor{DarkRed}{rgb}{0.55,0.00,0.00}
\definecolor{DarkGreen}{rgb}{0.00,0.55,0.00}
\definecolor{Purple}{rgb}{0.5,0.0,0.5}
\definecolor{Bittersweet}{rgb}{1.0,0.44,0.37}
\title{High level implementation of geometric multigrid solvers for finite element problems: applications in atmospheric modelling}

\ifbool{PREPRINT}{ 
\author[$\dagger$,1]{Lawrence Mitchell}
\author[*,2]{Eike Hermann M\"uller}
\affil[1]{Department of Computing and Department of Mathematics, Imperial College, South Kensington Campus, London SW7 2AZ}
\affil[$\dagger$]{Email: \texttt{lawrence.mitchell@imperial.ac.uk}}
\affil[2]{Department of Mathematical Sciences, University of Bath, Bath BA2 7AY, United Kingdom}
\affil[*]{Email: \texttt{e.mueller@bath.ac.uk}}

}{ 
\author[imperial]{Lawrence~Mitchell\fnref{fn2}}
\ead{lawrence.mitchell@imperial.ac.uk}
\fntext[fn2]{email: \texttt{lawrence.mitchell@imperial.ac.uk}}
\author[bath]{Eike~Hermann~M\"{u}ller\corref{cor1}\fnref{fn1}}
\ead{e.mueller@bath.ac.uk}
\cortext[cor1]{Corresponding author}
\fntext[fn1]{email: \texttt{e.mueller@bath.ac.uk}}
\address[imperial]{Department of Computing and Department of Mathematics, Imperial College, South Kensington Campus, London SW7 2AZ}
\address[bath]{Department of Mathematical Sciences, University of Bath, Claverton Down, Bath BA2 7AY, United Kingdom}
} 

\date{\today}
\begin{document}
\ifbool{PREPRINT}{ 
\maketitle
}{} 
\begin{abstract}
The implementation of efficient multigrid preconditioners for elliptic partial differential equations (PDEs) is a challenge due to the complexity of the resulting algorithms and corresponding computer code. 
For sophisticated (mixed) finite element discretisations on unstructured grids an efficient implementation can be very time consuming and requires the programmer to have in-depth knowledge of the mathematical theory, parallel computing and optimisation techniques on manycore CPUs.
In this paper we show how the development of bespoke multigrid preconditioners can be simplified significantly by using a framework which allows the expression of the each component of the algorithm at the correct abstraction level. Our approach (1) allows the expression of the finite element problem in a language which is close to the mathematical formulation of the problem, (2) guarantees the automatic generation and efficient execution of parallel optimised low-level computer code and (3) is flexible enough to support different abstraction levels and give the programmer control over details of the preconditioner. 
We use the composable abstractions of the Firedrake/PyOP2 package to demonstrate the efficiency of this approach for the solution of strongly anisotropic PDEs in atmospheric modelling. The weak formulation of the PDE is expressed in Unified Form Language (UFL) and the lower PyOP2 abstraction layer allows the manual design of computational kernels for a bespoke geometric multigrid preconditioner. We compare the performance of this preconditioner to a single-level method and hypre's BoomerAMG algorithm. The Firedrake/PyOP2 code is inherently parallel and we present a detailed performance analysis for a single node (24 cores) on the ARCHER supercomputer. Our implementation utilises a significant fraction of the available memory bandwidth and shows very good weak scaling on up to 6,144 compute cores.
\end{abstract}
\ifbool{PREPRINT}{ 
\textbf{keywords}:
\newcommand{\sep}{,}
}{ 
\begin{keyword}
} 
geometric multigrid\sep\ atmospheric modelling\sep\ preconditioner\sep\ (mixed) finite elements\sep\ domain-specific compilers
\ifbool{PREPRINT}{ 
\\[1ex]
}{ 
\end{keyword}
} 
\maketitle
\section{Introduction}
Efficient solvers for partial differential equations (PDEs) are
required in all areas of science and engineering.  The design and
implementation of these solvers requires a wide set of skills,
including, but not limited to: knowledge of the system being
simulated; creation and implementation of appropriate numerical
schemes; analysis of the resulting linear and nonlinear operators; and
efficient low-level implementation of the chosen schemes.  It is
therefore rare that a single individual possesses the full range of
skills to successfully deliver both an algorithmically and
computationally efficient solver on their own.

To address this multi-disciplinary problem requires software
frameworks that enable scientists with complementary skills and
specialisations to collaborate \emph{without} each one of them
requiring full knowledge of the system.  This can be achieved by
carefully designing composable interfaces that capture the natural
abstraction at each level of the simulation stack.

An important application is the simulation of fluid systems which are
described by the Navier Stokes equations.
In contrast to other mixed finite element methods such as the Taylor-Hood
element, discretisations based on mimetic finite element methods (see
e.g.~\cite{Brezzi1987,Cotter2012,Cotter2014}) exactly preserve certain
physical properties of the system under study. In particular, the divergence theorems $\nabla\times(\nabla \phi)=0$ and $\nabla\cdot(\nabla\times\vec{u})=0$ hold exactly for the \textit{discretised} operators; this is not the case for some other mixed methods such as the Taylor-Hood element. Mimetic methods
reproduce the
favourable properties of the C-grid staggering, but allow the use of
higher order discretisations and non-orthogonal grids.  This is
particularly important in global atmospheric modelling applications,
where the ``pole problem'' of standard longitude-latitude grids
introduces artificial time step limitations near the pole and
restricts the parallel scalability of the
code~\cite{Staniforth2012}. While mimetic finite element schemes are
very popular in applications such as numerical climate- and weather-
prediction, their efficient implementation poses significant
challenges. To deliver forecasts under tight operational timescales,
solvers have to be algorithmically robust, make optimal use of modern
manycore chip architectures and scale to large node counts on
distributed memory clusters.

Elliptic PDEs arise in implicit time stepping methods of the
atmospheric equations of motion and their solution often forms the
bottleneck of the full model code. Since the Earth's atmosphere is
represented by a thin spherical shell, the resulting discrete system is
highly anisotropic. This requires the use of bespoke
preconditioners, in particular the tensor-product multigrid approach in~\cite{BoermHiptmair1999} has turned out to algorithmically efficient~\cite{Mueller2014a,Mueller2014b,Dedner2015}.
The implementation of multigrid
algorithms for low-order finite difference discretisations on
structured grids is straightforward but this is not the case for
sophisticated finite element discretisations on more complex
geometries for several reasons: Since the HDiv (velocity) mass matrix
is not diagonal, the iterative solver algorithm is significantly more
involved than the corresponding finite difference version. In the
finite element case the innermost compute kernels can not be expressed
as simple stencil applications and - to achieve optimal performance on
a particular architecture - the loop nest has to be optimised, bearing
in mind hardware specific properties such as the cache layout. In a
distributed memory setting halo exchanges and overlapping of
computation and communication require a careful partitioning of the
grid. The Firedrake/PyOP2~\cite{Rathgeber2012,Rathgeber2015} framework allows
the automatic assembly of
finite element operators from their weak formulation and the
expression of the algorithm at a high abstraction level. Architecture
dependent optimised low-level C-code is automatically generated and
executed with just-in-time compilation techniques.

In this paper, we discuss the extension of the Firedrake framework
to support the development of geometric multigrid
solvers for finite element problems.  We illustrate its use, and
investigate the performance of the resulting method, in the
development of a geometric multigrid solver for a mimetic finite
element discretisation of the atmospheric equations of motion.  This
is a challenging problem and an excellent test case for the developed abstractions since both the implementation of the underlying finite element scheme and the design of an efficient multigrid method are non-trivial.  We
show how our approach simplifies the implementation of the solver, by
cleanly separating the different aspects of the model. In particular we
demonstrate how the chosen abstractions allow the easy implementation 
of a column-local matrix representation which is crucial for the
block-Jacobi smoother. This particular smoother is algorithmically optimal
for strongly anisotropic problems and a key ingredient of the tensor-product
multigrid algorithm in~\cite{BoermHiptmair1999}. A careful performance
analysis confirms that our implementation is efficient in the sense that it
uses a significant fraction of the system's peak performance for bandwidth-bound applications.

An alternative and very successful approach to the implementation of
finite element solvers is the use of templated C++
code~\cite{Bangerth2007,Bastian2010,Dedner2010}. This allows the user
full control over all components of the algorithm and multigrid
solvers for finite element discretisations have been implemented for
example in DUNE~\cite{Bastian2012} and deal.II~\cite{Kanschat2015}.
However, the key computational kernels (such as the local application
of the operator in weak form) have to be written by hand and any
optimisation is limited by the capabilities of the available
compiler. In contrast, frameworks like FEniCS~\cite{Logg2012} and
Firedrake use domain-specific compilers to carry out optimisations
that are infeasible for general purpose compilers to perform on a
low-level representation of the same algorithm~\cite{Luporini2016}.
Compared to FEniCS, where expressing non-finite element operations
requires the programmer to explicitly manage all parallelism and mesh
iteration, one of the advantages of our approach is the
straightforward implementation of any local operations as
computational kernels in PyOP2~\cite{Rathgeber2012}; this is crucial
for the preconditioners considered in this work.

The paper is structured as follows.  In
section~\ref{sec:FiredrakePyOP2}, we give a brief overview of the
Firedrake software framework.  Section~\ref{sec:multigrid-operators}
lays out the mathematical abstractions of the multigrid method, and
how we organise the software abstractions around them.  The multigrid
method for our model problem, an atmospheric gravity wave, is
discussed in section~\ref{sec:Multigrid3d}.  We characterise the
performance of the resulting scheme in section~\ref{sec:Results} and
conclude in section~\ref{sec:Conclusion}.
\ifbool{PREPRINT}{ 
Some more technical aspects are discussed in the appendices where we derive
the equations for linear gravity wave propagation, give the parameters of 
the PETSc fieldsplit solver and provide a detailed breakdown of the solver
setup times.
}{}
\paragraph{Main achievements}
We demonstrate the performance and parallel scalability of the solver
on the ARCHER supercomputer and compare our geometric multigrid
implementation to a matrix-explicit implementation based on the PETSc
library~\cite{petsc-efficient,petsc-user-ref}. In the latter case we
use the BoomerAMG~\cite{Henson2002} preconditioner from the hypre
suite~\cite{FalgoutYang2002} to solve the DG-pressure system.  We show
that the performance of the solver for the low order discretisations
treated in the paper is memory bound, and quantify the absolute
performance of the most computationally intensive kernels in our
solver algorithm by a detailed analysis of memory traffic.  We find
that the computationally most expensive kernels utilise a significant
fraction of the peak memory bandwidth.

\section{Firedrake/PyOP2: abstractions for finite element methods}\label{sec:FiredrakePyOP2}

Firedrake~\cite{Rathgeber2015} is a Python system for the solution of
partial differential equations by the finite element method.  It
builds on the abstractions introduced in the FEniCS
project~\cite{Logg2012,Logg2010} to present a high-level, automated,
problem solving environment.  Firedrake enforces a strong separation
of concerns between \emph{employing} the finite element method, the
\emph{implementation} of the local discretisation of the mathematical
operators, and their parallel execution over a mesh.  The execution of
kernels over the mesh is carried out using an iteration abstraction
layer, PyOP2~\cite{Rathgeber2012}.  This layer is explicitly exposed
to the model developer, and allows them to write and execute
\emph{custom} kernels over the mesh.  The critical observation is that
most operations that fall only slightly outside the finite element
abstraction may still be formulated as the execution of a \emph{local}
operation over some set of mesh entities, these can be expressed as a
PyOP2 \emph{parallel loop}.  This separate abstraction layer allows
the user to worry about the local operations: parallelisation is
carried out automatically by PyOP2 exactly as it is for finite element
kernels in Firedrake itself.  More details of the interaction between
Firedrake and PyOP2 can be found in~\cite[Section 4]{Rathgeber2015}.

\section{Multigrid in Firedrake}
\label{sec:multigrid-operators}

Multigrid methods~\cite{Brandt2011} are an algorithmically optimal approach
to solving many PDEs, especially those involving elliptic operators. 
They rely on a hierarchy of scales to cheaply
and efficiently compute properties of the system \emph{at the
  appropriate scale}: modes that vary slowly in space can be accurately
represented using
only a few degrees of freedom and are thus best solved for on coarse
grids, whereas fast modes need high spatial accuracy.  Here we briefly
provide an overview of the mathematical operations in terms of a
two-grid setup, and then describe our implementation in Firedrake.

Let $V_c$ be the approximation space on the ``coarse'' grid and $V_f$
the space on the fine grid, these need not necessarily be nested,
although the implementation is simplified if they are.  A complete
multigrid cycle is built from a few basic operators.  In
algorithm~\ref{alg:two-grid} we show the form of \emph{correction
  scheme} multigrid (so called because on the coarse grid we solve for
the correction to the fine grid equation) on two levels, given a
linear system $A \vec{u} = \vec{b}$.
\begin{algorithm}
\caption{Two-grid cycle}
\label{alg:two-grid}
  \centering
  \begin{algorithmic}[1]
    \STATE $\vec{u}_f \leftarrow \mathcal{S}_f(A_f, \vec{u}_f,
      \vec{b}_f$) \COMMENT{Presmooth (M1)}
    \STATE{$\vec{b}_c \leftarrow \mathcal{R}_c^f(\vec{b}_f -
      A_f\vec{u}_f)$} \COMMENT{Restrict residual (M2)}
    \STATE{$\vec{\delta u}_c \leftarrow A_c^{-1} \vec{b}_c$}
    \COMMENT{Solve for coarse correction (M3)}
    \STATE{$\vec{u}_f \leftarrow \vec{u}_f + \mathcal{P}_c^f(\vec{\delta
      u}_c)$} \COMMENT{Prolong correction (M4)}
    \STATE{$\vec{u}_f \leftarrow \mathcal{S}_f(A_f, \vec{u}_f,
      \vec{b}_f$)} \COMMENT{Postsmooth (M5)}
  \end{algorithmic}
\end{algorithm}
The extension to multiple levels recursively applies this two-level
cycle to compute $A_c^{-1} \vec{b_c}$.  This setup suggests that we
need to provide facilities for computing restrictions and
prolongations.  We also need the ability to compute coarse grid
operators, and we need to be able to apply smoothers (effectively some
form of linear solver).  In listing~\ref{lst:two-grid}, we show how
this abstract framework translates into our implementation in
Firedrake for a simple two-dimensional example.
We are able to exploit the existing facilities for the vast majority
of the implementation, we merely need a few extensions to deal with
hierarchies of meshes and transferring between them, which we discuss
below.

\begin{lstlisting}[language={[firedrake]{python}}, label=lst:two-grid, caption={Two-grid cycle in Firedrake for solving $\langle\nabla v\cdot \nabla u\rangle+\langle vu\rangle=\langle v f\rangle$ (the weak form of the sign-positive Helmholtz problem $-\Delta u + u = f$) with a piecewise linear $P_1$ discretisation on an icosahedral spherical mesh}]
from firedrake import *
# Construct mesh hierarchy ...
coarse = UnitIcosahedralSphereMesh(2)
mh = MeshHierarchy(coarse, 1)
# ... and corresponding function space hierarchy based on piecewise linear elements
Vh = FunctionSpaceHierarchy(mh, "CG", 1)
V = Vh[-1]

u = TrialFunction(V)
v = TestFunction(V)
# Define forms for A and b
a = (dot(grad(u), grad(v)) + u*v)*dx
x = SpatialCoordinate(mh[-1])
L = exp(-0.5*abs(x)**2/(0.5*0.5))*v*dx

# Solution on fine grid
uf = Function(V)
# Residual
R = L - action(a, uf)
Af = assemble(a)
bf = assemble(L)
# Pre-smooth (M1) with two point-Jacobi iterations
solve(Af, uf, bf, solver_parameters={'ksp_type': 'richardson',
                                     'ksp_max_it': 2,
                                     'ksp_convergence_test': 'skip',
                                     'ksp_initial_guess_nonzero': True,
                                     'pc_type': 'jacobi'})

duf = Function(V)
Vc = Vh[0]
uc = Function(Vc)
bc = Function(Vc)
# Restrict residual (M2)
restrict(assemble(R), bc)
Ac = assemble(coarsen_form(a))
# Exact coarse solve (M3)
solve(Ac, uc, bc, solver_parameters={'pc_type': 'lu',
                                     'ksp_type': 'preonly'})
# Prolongate correction (M4)
prolong(uc, duf)
uf += duf
# Post-smooth (M5) with three point-Jacobi iterations
solve(Af, uf, bf, solver_parameters={'ksp_type': 'richardson',
                                     'ksp_max_it': 3,
                                     'pc_type': 'jacobi',
                                     'ksp_convergence_test': 'skip',
                                     'ksp_initial_guess_nonzero': True})

\end{lstlisting}

\subsection{Grid hierarchies}
\label{sec:grid-hierarchies}
It is clear that the first object we will need is a hierarchy of grids
(or meshes, in Firedrake's parlance).  This will encapsulate the
relationship between the refined grids (providing information on the
fine grid cells corresponding to coarse grid cells).  These are
provided by the Firedrake \texttt{MeshHierarchy} object.  Similarly,
we shall need to represent discrete solution spaces.  Firedrake uses
\texttt{FunctionSpace} objects for this, and we extend these with a
\texttt{FunctionSpaceHierarchy}.  As with the \texttt{MeshHierarchy}
this encapsulates the relationship between function spaces on related
grids (allowing us to determine the degrees of freedom in the fine
grid that are related to a coarse grid cell).  We note that in this
work, we only treat hierarchically refined grids.  This does not
affect the abstractions we discuss, although it does simplify some of
the implementation.

\subsection{Restriction and prolongation}
\label{sec:grid-transfer}
Nested finite element spaces admit particularly simple implementation
of restriction and prolongation.  To compute the restriction operator
we use FIAT~\cite{Kirby2004} to evaluate the coarse basis at the node
points on the fine grid.  This allows us to express the coarse cell
basis functions in terms of linear combinations of fine cell basis
functions.  Since the mesh is regularly refined, we need only do this once
and can use the same weighting for all cells.  The restriction
operator can then be expressed simply by applying this combination
kernel to a given residual using a PyOP2 parallel loop (exposed as
\texttt{restrict} in the Firedrake interface).  In the same way we
compute the interpolation from $V_c$ into $V_f$, which is just the
natural embedding, using FIAT, and apply the kernel over the mesh with
PyOP2.

\subsection{Coarse grid operators}
\label{sec:coarse-grid-oper}
Forming the coarse grid operators is straightforward using the
existing facilities of Firedrake and UFL~\cite{Alnaes2014}.
Rediscretised operators are readily available simply by taking the UFL
expression for the fine grid operator and assembling it on the coarse
grid (achieved using the \texttt{coarsen\_form} operation in
listing~\ref{lst:two-grid}).  These rediscretised operators have
minimal stencil.  In addition, it is also straightforward to provide
simpler operators (perhaps throwing away couplings that do not
contribute on coarse grids) by explicitly defining the operator
symbolically on the appropriate coarse level and using it in the
smoother.

\subsection{Smoothers}
\label{sec:smoothers}
Firedrake uses PETSc to provide solvers for linear systems.  As such, for assembled matrices, we can use as a smoother any linear
solver that PETSc makes available.  In listing~\ref{lst:two-grid}, for
example, we use two Jacobi-preconditioned Richardson iterations as a
pre-smoother, solve the coarse problem exactly with LU and then use
three preconditioned Richardson iterations as a post-smoother.
Naturally, we are free to implement our own smoothers instead, perhaps we
do not have an assembled matrix and therefore cannot use ``black-box''
smoothers.  Indeed, the key ingredient to achieve optimal performance is the
use of the correct smoother.
This is of particular importance in the tensor-product
multigrid scheme we discuss in the rest of the paper, since an
assembled operator is not available and there is structure in the
problem (the strong vertical anisotropy) we wish to exploit in the smoothers.

\section{Tensor-product multigrid for (mimetic-) mixed finite
  elements}\label{sec:Multigrid3d}
To illustrate the power of these abstractions we consider an important model
system for meteorological applications:
the equations for linear gravity wave propagation in the global
atmosphere. The corresponding system of PDEs is discretised with a
mixed finite element discretisation. The problem is solved in a thin
spherical shell which represents the Earth's atmosphere. The thickness
of the atmosphere is several orders smaller than the radius of the
earth.  This flatness of the domain is typical for applications in
atmospheric modelling and introduces a strong grid-aligned
anisotropy. As we shall see, for a mixed finite element discretisation this
problem is significantly more complex than the simple example shown in
listing~\ref{lst:two-grid}. We will therefore use it to both
characterise the performance of our multigrid implementation, and validate
that we have exposed the correct abstractions. 
For the linear solver to converge rapidly, the anisotropy
has to be treated correctly in the preconditioner and the main challenge is
the implementation of an optimal smoother. The
PyOP2 abstraction level allows the expression of this smoother in terms of
tailored data structures and low-level kernels which implement the
line-relaxation method that is key to exploit the vertical anisotropy.

Since the PDE we are solving is elliptic, one option is to employ an
algebraic multigrid (AMG) preconditioner.  On highly anisotropic
domains, one must take special care in constructing the coarse grid
operators and smoothers to achieve mesh independence.  See for
example~\cite{Tobin2015}, where the authors use smoothed-aggregation
AMG to precondition the velocity block when solving the nonlinear
Stokes equations in the context of ice-sheet dynamics.  To account for
the strong anisotropy, the aggregation strategy is adapted to maintain
the column structure of the degrees of freedom, and a smoother based
on incomplete factorisations is used. There are a number of reasons
why an AMG approach might not always be desirable. The AMG
preconditioner can only use Galerkin coarse grid operators, and so the
coarse grid stencil will be large.  More importantly, unless special
care is taken, the coarsening operation will not necessarily obey the
anisotropy inherent in the problem, perhaps leading to suboptimal
algorithmic performance.  Finally, we found in~\cite{Mueller2014a}
that a bespoke preconditioner, based on the tensor-product multigrid
of~\cite{BoermHiptmair1999}, is superior to AMG based methods if it is
applied to a simplified model equation discretised with the finite
volume method; indeed the geometric multigrid approach turned out to
be about $10\times$ faster than black-box AMG implementations from the
DUNE~\cite{Blatt2007} and hypre~\cite{FalgoutYang2002} libraries.

Moreover the tensor-product preconditioner can be shown to be optimal for
grid-aligned anisotropies. In a recent paper~\cite{Dedner2015} we have
also demonstrated numerically that the method works well for
atmospheric applications under slightly more general
circumstances. These encouraging results motivate us to study the
geometric multigrid implementation reported in this work.

\subsection{Mathematical formulation and mixed finite element discretisation}
The following linear gravity wave problem for pressure $p$, velocity $\vec{u}$ and buoyancy $b$ can be obtained by linearising the full Navier-Stokes equations for large scale atmospheric flow%
\ifbool{PREPRINT}{
 (see appendix~\ref{sec:ModelEquationDerivation} for a derivation):
}{
:
} 
\begin{xalignat}{3}
   \ddt{\vec{u}} &= \nabla p + b \zhat, &
   \ddt{p} &= -c^2 \nabla\cdot \vec{u}, &
   \ddt{b} &= -N^2\vec{u}\cdot\zhat.
\label{eqn:ContinuousEquations}
\end{xalignat}
For simplicity we assume that both the speed of sound
$c\approx 300ms^{-1}$ and the buoyancy frequency $N\approx 0.01s^{-1}$
are constant and enforce the (strong) boundary condition
\begin{equation}
 \vec{u}\cdot\vec{n}=0\label{eqn:BoundaryCondition}
\end{equation}
at the upper and lower boundary of the atmosphere. The domain $\Omega$
can be expressed as a tensor-product
$\Omega = S^2(R)\times
[0,H]$ where $S^2(R)$ is the
two-dimensional surface of a sphere with radius $R$. As described
above, the domain is very flat, i.e.
$H \ll R$. To construct
function spaces for mimetic finite element discretisations, consider
the following de Rham complexes in one-, two- and three dimensions:
\begin{xalignat}{3}
  \Vspace_0 \overset{\partial_z}{\rightarrow}\Vspace_1,&
  &\Uspace_0 \overset{\nabla^{\perp}}{\rightarrow}
\Uspace_1 \overset{\nabla\cdot}{\rightarrow}\Uspace_2, &
  &\Wspace_0 \overset{\nabla}{\rightarrow} \Wspace_1 \overset{\nabla\times}{\rightarrow} \Wspace_2\overset{\nabla\cdot}{\rightarrow} \Wspace_3.
  \label{eqn:all_spaces}
\end{xalignat}
We seek a solution to Eq.~\ref{eqn:ContinuousEquations} with
\begin{xalignat}{3}
  \vec{u} &\in \Wspace_2^0 = \Wspace_2^h \oplus \Wspace_2^{0,z}&
  b &\in \Wspace_b, &
  p &\in \Wspace_3,
  \label{eqn:FunctionSpaces}
\end{xalignat}
where $\Wspace_2^0$ is the subspace of $\Wspace_2$ whose normal
component vanishes on the boundary of the domain.  $\Wspace_2^h =
\Hdiv(\Uspace_2\otimes\Vspace_0)$ and $\Wspace_2^{0,z} =
\Hdiv(\Uspace_1\otimes\Vspace_1)$ are respectively the ``horizontal''
and ``vertical'' parts of $\Wspace_2^0$. The remaining spaces are $\Wspace_b =
\Uspace_1\otimes\Vspace_1$ and $\Wspace_3 =
\Uspace_2 \otimes \Vspace_1$. This choice of spaces for $\vec{u}$ and
$b$ is analogous to the Charney-Phillips staggering.
We refer the reader to~\cite{McRae2015} for the implementation and
further description of tensor-product spaces in Firedrake.

It is worth highlighting here that the decomposition of the velocity space
into a horizontal
($\Wspace_2^h$) and vertical ($\Wspace_2^{0,z}$) component is very
important for the construction of the tensor-product multigrid
preconditioner described below.

We discretise in time using an implicit scheme which is a special case
of the Crank-Nicholson method~\cite{CrankNicolson1996},
resulting in the following weak system for the increments
$\delta \vec{u}\in \Wspace_2^0$, $\delta b\in\Wspace_b$ and
$\delta p\in\Wspace_3$
\begin{equation}
 \begin{aligned}
  \langle\vec{w},\delta\vec{u}\rangle
  - \frac{\Delta t}{2}\langle\nabla\cdot\vec{w},\delta p\rangle
  - \frac{\Delta t}{2}\langle\vec{w},\delta b\zhat\rangle
  &= \Delta t \langle\nabla\cdot\vec{w},p_0\rangle
  + \Delta t \langle\vec{w},b_0\zhat\rangle
  \equiv \langle \vec{w},\vec{r}_u\rangle &\forall \vec{w}\in\Wspace_2^0
 \\[0ex]
  \langle\phi,\delta p\rangle
  +\frac{\Delta t}{2}c^2\langle\phi,\nabla\cdot\delta\vec{u}\rangle
  &= -\Delta tc^2\langle\phi,\nabla\cdot\vec{u}_0\rangle
  \equiv \langle\phi,r_\phi\rangle &\forall \phi\in\Wspace_3
  \\[0ex]
  \langle\gamma,\delta b\rangle
  + \frac{\Delta t }{2}N^2\langle\gamma,\delta\vec{u}\cdot\zhat\rangle
  &= -\Delta t N^2\langle\gamma,\vec{u}_0\cdot\zhat\rangle
  \equiv \langle \gamma,r_b\rangle &\forall \gamma\in\Wspace_b
\label{eqn:Increments}
 \end{aligned}
\end{equation}
where $\hat{\vec{z}}$ is the normal vector in the vertical direction
and $\vec{u}_0$, $p_0$ and $b_0$ are the known fields are the previous
time step.  In this work we consider two choices for the
three-dimensional complex in Eq.~\ref{eqn:all_spaces} introduced
in~\cite{Cotter2012} and summarised in
Tab.~\ref{tab:FiniteElements}. In the following they are referred to
as ``Lowest Order'' (LO) and ``Next-to-Lowest Order'' (NLO).
\begin{table}
 \begin{center}
 \begin{tabular}{lcccc}
  & $\Vspace_0$ & $\Vspace_1$ & $\Uspace_1$ & $\Uspace_2$\\
  \hline
  Lowest Order (LO) & $P_1$ & $DG_0$ & $RT_0$ & $DG_0$\\
  Next-to-Lowest Order (NLO) & $P_2$ & $DG_1$ & $BDFM_1$ & $DG_1$
 \end{tabular}
 \caption{Finite element discretisations used in this work. $P_n$ is the continuous polynomial element of degree $n$, $DG_n$ the corresponding discontinuous elements of the same degree. $RT_0$ is the lowest order Raviart-Thomas element \cite{Raviart1977} and $BDFM_1$ is the element described in \cite{Brezzi1987}.}
 \label{tab:FiniteElements}
 \end{center}
\end{table}
Discretising, we obtain a system of linear equations for the
dof-vectors $\vec{U}$, $\vec{P}$ and $\vec{B}$:
\begin{equation}
\begin{pmatrix}
  M_2 & 
    -\frac{\Delta t}{2}D^T & 
    -\frac{\Delta t}{2}Q\\[1ex]
  \frac{\Delta t}{2}c^2D & M_3 & 0\\[1ex]
  \frac{\Delta t}{2}N^2Q^T & 0 & M_b
\end{pmatrix}
\begin{pmatrix}
  \vec{U}\\[1ex]\vec{P}\\[1ex]\vec{B}
\end{pmatrix}
=
\begin{pmatrix}
M_2\vec{R}_u\\[1ex]
M_3\vec{R}_p\\[1ex]
M_b\vec{R}_b
\end{pmatrix}
\label{eqn:Equations3x3}
\end{equation}
Note that since $\Wspace_2^0=\Wspace_2^h\oplus\Wspace_2^{0,z}$, the velocity mass matrix can be written as the tensorial sum of a horizontal and a vertical mass matrix
\begin{equation}
  M_2 = M_2^h \oplus M_2^z.
\end{equation}
Similarly the weak derivative can be expressed as $D=D_h\oplus D_z$.

In the absence of orography, buoyancy can be eliminated point wise from Eq.~\ref{eqn:Equations3x3} to obtain a block- $2\times 2$ system of equations for velocity and pressure only\footnote{Note that the continuous version of the last equation in (\ref{eqn:Equations3x3}), $\delta b+\tfrac{\Delta t}{2} N^2 \delta\vec{u}\cdot \hat{\vec{z}}+\Delta t N^2 \vec{u}_0\cdot\hat{\vec{z}}=0$ is true also in the presence of orography, and an alternative approach (leading to a different discrete system) would be to eliminate orography from the continuous equations before discretising. However, a pointwise elimination of buoyancy from the \textit{discretised} equations is only possibly in the absence of orography when (\ref{eqn:Equations3x3}) holds strongly and not just weakly.},
\begin{equation}
  A\begin{pmatrix}\vec{U}\\[1ex]\vec{P}\end{pmatrix}
  \equiv
  \begin{pmatrix}
   \tilde{M}_2 & -\frac{\Delta t}{2}D^T  \\[1ex]
   \frac{\Delta t}{2}c^2 D & M_3
  \end{pmatrix}
  \begin{pmatrix}\vec{U}\\[1ex]\vec{P}\end{pmatrix}
 =
\begin{pmatrix}M_2\tilde{\vec{R}}_u\\[1ex]M_3\vec{R}_p\end{pmatrix}
\qquad\text{with}\quad 
\begin{matrix}
\tilde{M}_2 = M_2^h \oplus (1+ \omega_N^2) M_2^z, \quad \omega_N \equiv \frac{\Delta t}{2}N
\\[1ex]
\tilde{\vec{R}}_u = \vec{R}_u+\frac{\Delta t}{2}M_2^{-1}Q\vec{R}_b.
\end{matrix}
\label{eqn:PressureVelocitySystem}
\end{equation}
It is important to note that, in contrast to lowest order finite volume discretisations on staggered grids, the velocity mass matrix $\tilde{M}_2$ is \textit{not} diagonal\footnote{It is, however, well conditioned, with a condition number of $\order(10)$ that is independent of the grid resolution.}. This prevents the standard treatment taken in many atmospheric modelling codes based on finite volume or finite difference discretisations (for example the semi-implicit dynamical cores of the Unified Model~\cite{Davies2005,Wood2014}), which eliminate velocity point wise to obtain a system for the pressure, which is solved iteratively before reconstructing the velocity. In our case we have to use an iterative solver for the full linear system in $\vec{U}$ and $\vec{P}$.

The condition number $\kappa$ of the linear system grows with the (vertical) acoustic Courant number $c\Delta t/\Delta z\gg 1$, and hence preconditioning is essential to achieve rapid convergence. Due to the flatness of the grid cells ($\Delta z \ll \Delta x$) the dominant terms in $\kappa$ are due to fast vertical sound waves. However, wave propagation can be treated approximately independently in each column of the grid (in a given time interval, a wave which propagates over $k$ grid cells in the vertical direction will only cover a distance of $k\cdot \Delta z/\Delta x\ll k$ cells in the horizontal). In the following we describe the construction of a preconditioner which uses this idea to reduce the condition number such that it only depends on the much smaller horizontal acoustic Courant number $\nuCFL=c\Delta t/\Delta x$ which is $\nuCFL=\order(10)$ in atmospheric models (larger values of $\nuCFL$ are excluded due to the explicit treatment of other processes such as advection). Under these conditions a tensor-product multigrid V-cycle with vertical line relaxation and horizontal grid coarsening will then lead to rapid convergence of the linear solver iteration.
\subsection{Schur-complement preconditioner}\label{sec:GeometricMGPreconditioner}
Formally the Schur-complement factorisation of the inverse of the block- $2\times 2$ matrix $A$ in Eq.~\ref{eqn:PressureVelocitySystem} is given by
\begin{equation}
  A^{-1} = 
\begin{pmatrix}
  1 & \frac{\Delta t}{2}\tilde{M}_2^{-1} D^T \\[1ex]
  0 & 1
\end{pmatrix}
\begin{pmatrix}
  \tilde{M}_2^{-1} & 0 \\[1ex] 0 & H^{-1}
\end{pmatrix}
\begin{pmatrix}
  1 & 0 \\[1ex]
  -\frac{\Delta t}{2}c^2D\tilde{M}_2^{-1} & 1
\end{pmatrix}\label{eqn:SchurComplement}
\end{equation}
with the elliptic and positive-definite ``Helmholtz'' operator
\begin{equation}
  H \equiv M_3 + \omega_c^2 D\tilde{M}_2^{-1} D^T
\qquad\text{where}\quad \omega_c \equiv \frac{\Delta t}{2}c.
  \label{eqn:HelmholtzOperator}
\end{equation}

This operator contains the dense inverse $\tilde{M}_2^{-1}$ of the
velocity mass matrix and is therefore also dense.
We obtain a preconditioner $P^{-1}\approx A^{-1}$ with a number of
approximation steps.
We replace the full inverses $\tilde{M}_2^{-1}$ in Eq.~\ref{eqn:SchurComplement} by a diagonal approximation $\tilde{M}_{2,\operatorname{inv}}$:
\begin{equation}
(\tilde{M}_{2,\operatorname{inv}})_{ij}=\delta_{ij}/(\tilde{M}_2)_{ii}.
\label{eqn:LumpedMass}
\end{equation}
Furthermore, since $\tilde{M}_2^{-1} = \left(M_2^{h}\right)^{-1}\oplus \frac{1}{1+\omega_N^2}\left(M_2^{z}\right)^{-1}$ and $D=D_h\oplus D_z$, the second order term in Eq.~\ref{eqn:HelmholtzOperator} can be written as 
\begin{equation}
  D\tilde{M}_2^{-1}D^T = D_h \left(M_2^h\right)^{-1} D_h^T \oplus \frac{1}{1+\omega_N^2}D_z \left(M_2^z\right)^{-1} D_z^T.
\end{equation}
Again, we replace full mass matrix inverses by their diagonal
approximations, $(M_2^h)^{-1}\mapsto M_{2,\operatorname{inv}}^h$,
$(M_2^z)^{-1}\mapsto M_{2,\operatorname{inv}}^z$.  Since the function
space $\Wspace_2^{0,z}$ is horizontally discontinuous, the matrix
$\tilde{M}_2^z$ has a block-diagonal structure.  It would therefore be
possible to use more sophisticated approximations for
$(\tilde{M}_{2}^z)^{-1}$ in each vertical column.  For example we
tried a sparse approximate inverse~\cite{Grote1997}, but found that
this leads to increased setup costs and worse overall performance.

With the diagonally lumped mass matrices we obtain the approximate Helmholtz operator
\begin{equation}
  \hat{H} = M_{3} + \omega_c^2\left(
  D_h M^h_{2,\text{inv}} D_h^T +
\frac{1}{1+\omega_N^2}D_z M^z_{2,\text{inv}} D_z^T\right)\approx H
\label{eqn:Preconditioner}
\end{equation}
which can be inverted iteratively to approximate $\hat{H}^{-1}\approx H^{-1}$ in Eq.~\ref{eqn:SchurComplement}.
In contrast to $H$, the operator $\hat{H}$ has a sparse structure but still contains the dominant couplings in the vertical direction.
Since the size of the derivatives $D_h$ and $D_z$ is proportional to the inverse grid spacings, the two terms in the brackets are of order $1/\Delta x^2$ and $1/\Delta z^2$ respectively (the pressure mass term $M_3$ is of order $1$). On the highly anisotropic grids considered here we have $\Delta z \ll \Delta x$, and hence the second term in the bracket is the dominant contribution. If we write $\Delta_h$ for the block-diagonal part of $D_h M^h_{2,\text{inv}} D_h^T$, the following block-diagonal operator differs from $\hat{H}$ by terms of $\order((\Delta z/\Delta x)^2)$:
\begin{equation}
  \hat{H}_z \equiv M_{3} + \omega_c^2\Delta_h+\frac{\omega_c^2}{1+\omega_N^2} D_z M^z_{2,\text{inv}} D_z^T=\hat{H}+\order((\Delta z/\Delta x)^2)
  \label{eqn:HHgeneralbanded}.
\end{equation}
The operator $\hat{H}_z$ is decoupled in the horizontal direction, and
can therefore be inverted independently in each column.  With a
suitable degree of freedom ordering this can be carried out using a
simple banded matrix solve.  To invert the operator $\hat{H}$, one can
now use a suitably preconditioned iterative method. One possibility
would be a Krylov iteration with block-Jacobi preconditioner, namely
\begin{equation}
  \vec{P} \mapsto \vec{P} + \omega\hat{H}_z^{-1}\left(\vec{P}-\hat{H}\vec{R}_{p}\right)\qquad\omega\in \mathbb{R}.
\label{eqn:BlockJacobi}
\end{equation}
This is similar to the approach taken in existing operational codes such as the Met Office Unified Model~\cite{Davies2005,Wood2014}. However, as we demonstrated in~\cite{Mueller2014a}, a much more efficient preconditioner is a tensor-product multigrid algorithm in which the grid is only coarsened in the horizontal direction and Eq.~\ref{eqn:BlockJacobi} is used as a smoother (we refer to this block-diagonal smoother as ``vertical line relaxation'' in the following). This is the method we use in this paper, and the numerical results in section~\ref{sec:Results} confirm the superiority of the method compared to a single-level preconditioner: the multigrid algorithm is about twice as fast. This was also observed in~\cite{Mueller2014a} and in numerical experiments we carried out on the ENDGame dynamical core. Since the inverse of $\hat{H}$ is only required in the preconditioner and does not have to be computed to very high accuracy, we found it is most efficient to simply apply one multigrid V-cycle with $\hat{H}$.
\subsubsection{Tensor-product multigrid algorithm}
\label{sec:tensor_product_multigrid}
The tensor-product multigrid algorithm which we use to solve the
system $\hat{H}\vec{P}=\vec{R}_{\vec{P}}$ was first described and
analysed in~\cite{BoermHiptmair1999} and proven robust for
grid-aligned anisotropies. In~\cite{Dedner2015} the analysis is
extended to $2+1$ dimensional grids and it is demonstrated
numerically that the algorithm is still efficient for approximately
grid aligned anisotropies in meteorological applications.

This can be interpreted in the context of the discussion at the end of session~\ref{sec:GeometricMGPreconditioner}: the
performance of the tensor-product multigrid algorithm
is given by the horizontal acoustic Courant number $\nuCFL\equiv c\Delta t/\Delta
x$; the much larger vertical Courant number $c\Delta t/\Delta z$ does not
appear
at leading order since sound propagation in each column is treated
exactly by the block-diagonal line relaxation smoother.

The tensor-product multigrid cycle fits naturally into the language of
section~\ref{sec:multigrid-operators}, with the following selections
for the operators.  Rather than full coarsening in all three
dimensions, we semi-coarsen in the horizontal direction, leaving
columns intact.  For the next-to-lowest-order function spaces, the first
coarsening step is a $p$-refinement from $DG_1$ into
$DG_0 \subset DG_1$, with the restriction and prolongation defined
using the natural embedding.  We use the vertical ``line relaxation''
operator of Eq.~\ref{eqn:BlockJacobi} for our multigrid smoother.  Finally we
note that the natural correlation length $\Lambda$ in units of the
horizontal grid spacing on the finest multigrid level is given by
$\Lambda=\nuCFL$. Since other components of the model (such as explicit
advection) limit the time step size to
$\Delta t\lesssim 10\Delta x/c$, this correlation length is
$\Lambda\lesssim 10$.  It is therefore sufficient to coarsen to the
grid until the grid spacing is larger then this length scale; the
coarse grid problem becomes well conditioned and can be solved with a
small number of smoother iterations.

\subsection{Implementation}
The function spaces $\Wspace_2^{0,z}$ and $\Wspace_3$ are discontinuous in
the horizontal direction. It is natural to order the degrees of
freedom such that all unknowns in a vertical column of the three
dimensional grid are stored consecutively in memory (see
Fig.~\ref{fig:bandedmatrix}). This storage format, used by
Firedrake, has good performance characteristics since the contiguous
column data can be directly addressed.

As a consequence, in Eq.~\ref{eqn:Preconditioner} and
Eq.~\ref{eqn:HHgeneralbanded} the matrices $M_3$, $\Delta_h$, $D_z$,
$M^z_{2,\operatorname{inv}}$ and $\hat{H}_z$ are columnwise
block-diagonal. Each block can be associated with a cell of the
two-dimensional grid that covers $S^2(R_{\operatorname{earth}})$ and
corresponds to a banded matrix which couples the unknowns in a
particular vertical column.  To apply the vertical line relaxation
smoother of Eq.~\ref{eqn:BlockJacobi} we explicitly assemble
$\hat{H}_z$ so that it is amenable to inversion using banded matrix
solves.  Although it is not possible to express the operations we
require using UFL -- they are not operations on variational forms --
it is still possible to write them as PyOP2 kernels that are executed
over the $S^2(R_{\operatorname{earth}})$ grid.  A key part of our
implementation are therefore linear algebra operations on columnwise
banded matrices.  We describe this algebra and its implementation in the
following section.

\subsubsection{Banded matrix algebra}\label{sec:BandedMatrixAlgebra}
\begin{figure}
 \begin{center}
   \begin{minipage}{0.48\linewidth}
     \includegraphics[width=0.9\linewidth]{\figdir/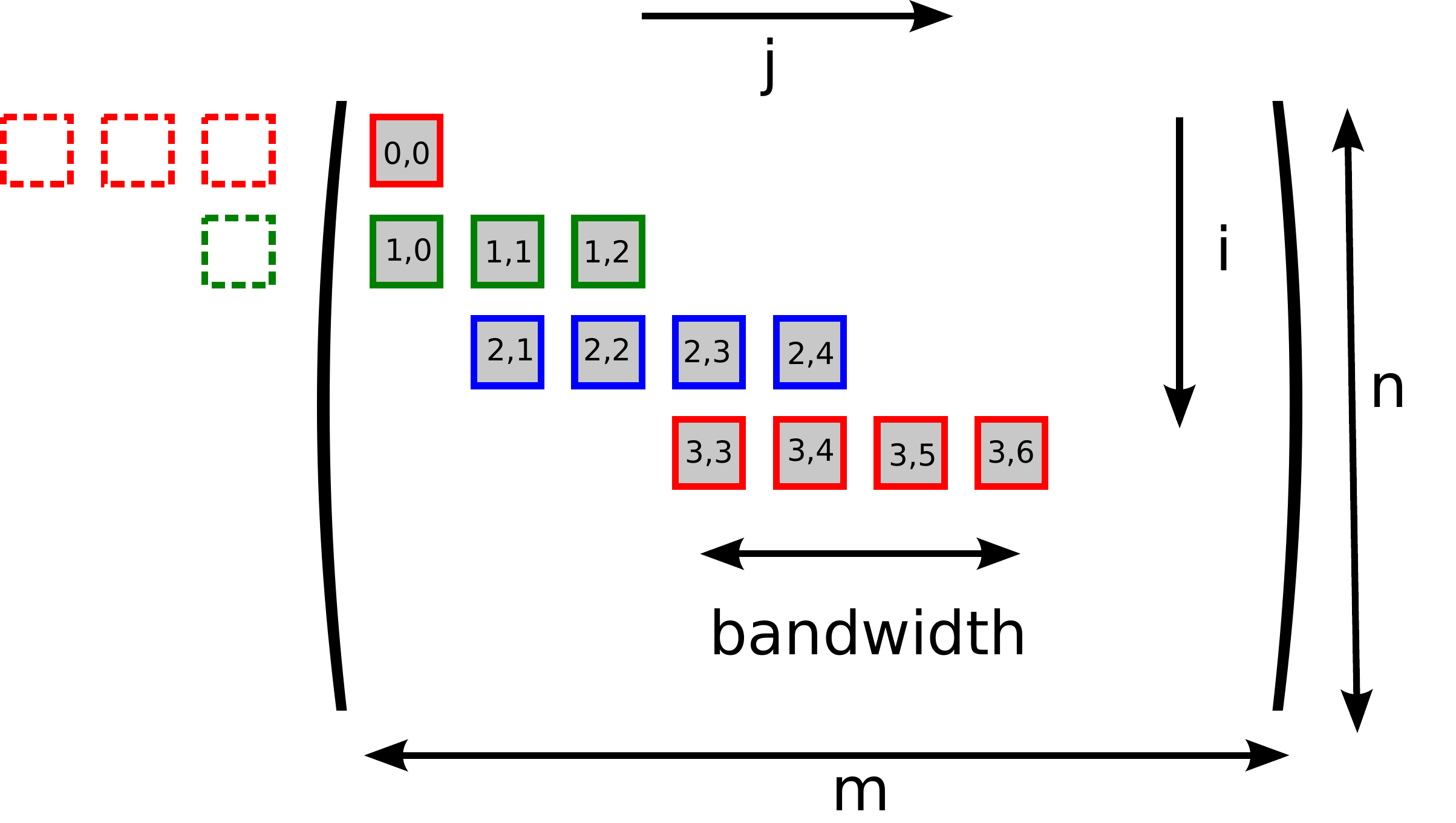}
   \end{minipage}
   \hfill
   \begin{minipage}{0.48\linewidth}
     \includegraphics[width=0.7\linewidth]{\figdir/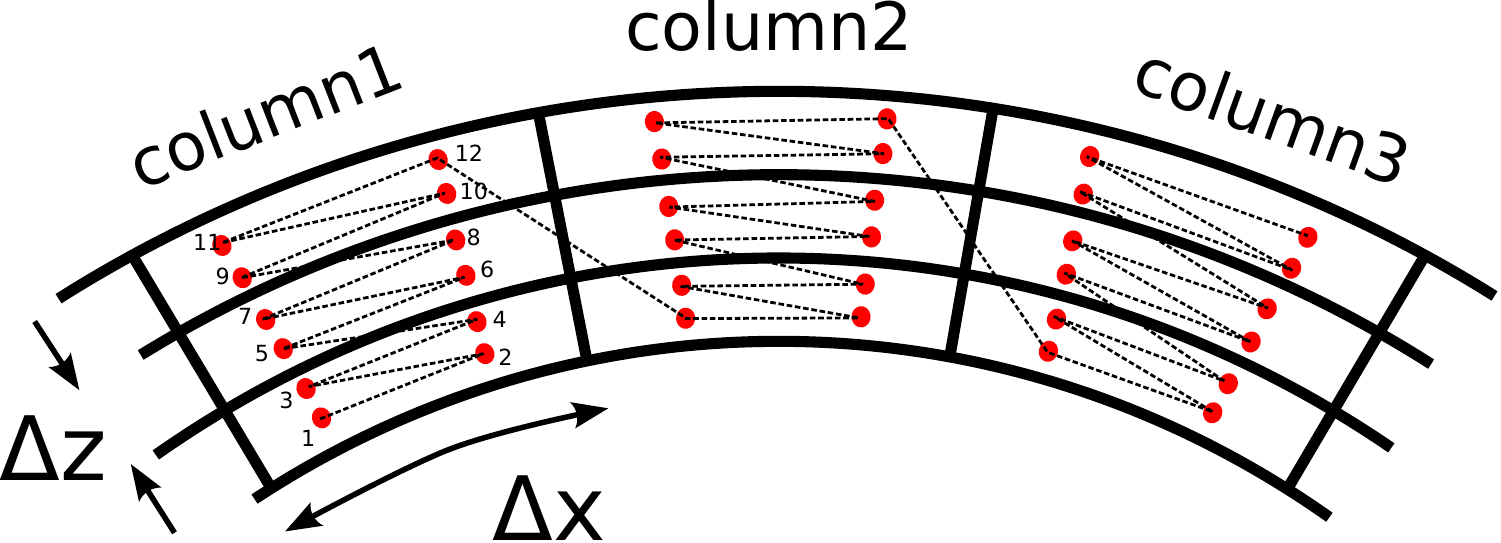}\\[8ex]
     \includegraphics[width=0.7\linewidth]{\figdir/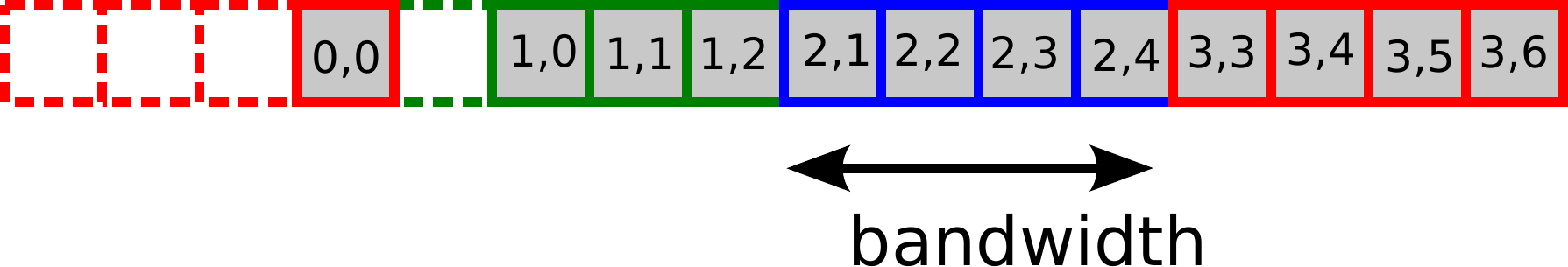}
   \end{minipage}
  \caption{
Columnwise ordering of degrees of freedom for discontinuous function spaces (upper right), generalised banded matrix (left) and sparse storage format (lower right) for $\alpha^{(\mathscr{L})}=2$, $\beta^{(\mathscr{L})}=1$, $\gamma_-^{(\mathscr{L})}=0$, $\gamma_+^{(\mathscr{L})}=3$ and bandwidth $4$.}
  \label{fig:bandedmatrix}
 \end{center}
\end{figure}

To represent the block-diagonal matrices we implemented a bespoke
banded matrix class which provides the necessary operations. In each
vertical column, given a UFL expression for the operator, we can
assemble the local matrix block. More generally, the class
implements the block-diagonal matrix representation $A[\mathscr{L}]$
of a linear operator $\mathscr{L}$ that maps between two horizontally
discontinuous spaces
\begin{equation}
  \mathscr{L}: \Wspace_{x} \rightarrow \Wspace_{y}
\end{equation}
In each vertical column $c$, the local matrix block $A^{(c)}[\mathscr{L}]$ is a generalised banded matrix:
\begin{equation}
  \left(A^{(c)}[\mathscr{L}]\right)_{ij} \begin{cases} \ne 0 & \text{for all}\quad i,j \;\; \text{with}\;\;
  -\gamma_i^{(\mathscr{L})} \le \alpha^{(\mathscr{L})}i-\beta^{(\mathscr{L})}j \le \gamma_+^{(\mathscr{L})}\\
  =0 & \text{otherwise}
  \end{cases}
\end{equation}
where $\gamma_{\pm}^{(\mathscr{L})}$, $\alpha^{(\mathscr{L})}$ and
$\beta^{(\mathscr{L})}$ (with
$\operatorname{gcd}(\alpha^{(\mathscr{L})},\beta^{(\mathscr{L})})=1$)
depend on the function spaces $\Wspace_x$ and $\Wspace_y$.  When
$\alpha^{(\mathscr{L})}=\beta^{(\mathscr{L})}=1$ (which occurs when
$\Wspace_x$ and $\Wspace_y$ have the same degree-of-freedom layout)
this reduces to the standard banded storage format.  In each
column, we then store the operator using a generalisation of the
LAPACK banded matrix format, see Fig.~\ref{fig:bandedmatrix}.

The Python class which implements $A[\mathscr{L}]$ provides the following functionality:
\begin{itemize}
  \item Assemble $\mathscr{L}$ from a given UFL form to obtain $A[\mathscr{L}]$
  \item Apply $A[\mathscr{L}]$ to a field-vector: $\Wspace_x\ni x\mapsto y=A[\mathscr{L}]x\in\Wspace_y$
  \item For two assembled linear operators $A[\mathscr{L}_1]$ and $A[\mathscr{L}_2]$, calculate $A[\alpha \mathscr{L}_1+\beta \mathscr{L}_2]=\alpha A[\mathscr{L}_1]+\beta A[\mathscr{L}_2]$ and $A[\mathscr{L}_1\cdot \mathscr{L}_2]=A[\mathscr{L}_1]\cdot A[\mathscr{L}_2]$.
  \item Apply the inverse of $A[\mathscr{L}]$ to a field-vector, i.e. solve the linear banded system $A[\mathscr{L}]x=y\in\Wspace_y$ for $x\in \Wspace_x$.
\end{itemize}
To solve the equation $A[\mathscr{L}]x=y$ in each column we use two
different approaches: at next-to-lowest order where the matrix bandwidth is
27, we use the LAPACK routine \verb+dgbtrf+ to precompute and store a
LU decomposition of the matrix $A[\mathscr{L}]$ and then employ the
corresponding method \verb+dgbtrs+ to solve the linear system based on
this LU decomposition. However, at lowest order, where the matrix is
tridiagonal, we found that a handwritten direct solve (Thomas
algorithm, see e.g.~\cite[\S 2.4]{Press2007}) is much more
efficient.

All operations can be carried out independently in each vertical
column. Based on this observation, the hierarchical architecture of
PyOP2/Firedrake is crucial for the implementation: we store the local
matrix blocks as PyOP2 dats on the horizontal grid cells. The
operations above can then be implemented as short PyOP2 kernels, which
are executed over the horizontal grid. An example is given in listing~\ref{lst:bandedkernel}, which shows the matrix-vector product $\vec{v}=A\vec{u}$. Note that the user only has to express the \emph{local} kernel as a short piece of C code, the PyOP2 system is responsible for executing this kernel over the grid.
\begin{lstlisting}[language={[firedrake]{python}}, label=lst:bandedkernel, caption={PyOP2 kernel for banded matrix-vector multiplication $\vec{v} = A\vec{u}$}]
# C-kernel code for Matrix-vector product v = A.u in one vertical column
kernel_code = '''void ax(double **A, double **u, double **v) {
  for (int i=0;i<n_row;++i) {                                   
    v[0][i] = 0.0;                                                     
    int j_m = (int) ceil((alpha*i-gamma_p)/beta);   
    int j_p = (int) floor((alpha*i+gamma_m)/beta);  
    for (int j=std::max(0,j_m);j<std::min(n_col,j_p+1);++j)
      v[0][i] += A[0][bandwidth*i+(j-j_m)] * u[j];
  }'''
# Execute PyOP2 kernel over grid
kernel = op2.Kernel(kernel_code,'ax',cpp=True)
op2.par_loop(kernel,hostmesh.cell_set, A(op2.READ,Vcell.cell_node_map()),
             u.dat(op2.READ,u.cell_node_map()),
             v.dat(op2.WRITE,u.cell_node_map()))
\end{lstlisting}
\subsection{PETSc/BoomerAMG based fieldsplit preconditioner}\label{sec:PETScFieldsplit}
To compare the performance of our bespoke implementation to an
existing solver package, we also solved the discretised equations
purely algebraically with PETSc~\cite{petsc-user-ref,petsc-efficient}
and preconditioned with hypre~\cite{FalgoutYang2002}.  PETSc's fieldsplit
preconditioner is used to algebraically form the Schur complement operator.
We use a diagonal approximation to the velocity mass inverse to form a
preconditioning operator for the Schur complement (corresponding
exactly with $\hat{H}$ in Eq.~\ref{eqn:Preconditioner}), everywhere
else the velocity mass matrix is inverted using an ILU-preconditioned
block-Jacobi iteration.  The
resulting elliptic problem in the pressure space is solved using the
BoomerAMG algebraic multigrid preconditioner from the hypre library.
To handle the strong vertical
anisotropy it was essential to replace the standard point-smoother by
an incomplete LU factorisation (Euclid). Since the vertical couplings
are the dominant terms, this is approximately the same as an exact
inversion of the block-diagonal operator which contains the vertical
couplings only.
\ifbool{PREPRINT}{
  Table~\ref{tab:fieldsplitoptions} in appendix~\ref{sec:fieldsplitoptions} details the full set of PETSc options
  used. %
}{
}
\section{Results}\label{sec:Results}
In all cases the linear system in Eq.~\ref{eqn:PressureVelocitySystem}
is solved with a suitably preconditioned, restarted GMRES($k=30$) iteration~\cite{Saad1986} to a
relative tolerance of $||\vec{r}||/||\vec{r}_0|| < 10^{-5}$, which is
typical in atmospheric modelling applications. Note that the system in
(\ref{eqn:Equations3x3}) is not positive definite (after all, it has
wave-like solutions), ruling out use of the more efficient Conjugate
Gradient method.  In the particular case of constant $c^2$ and $N^2$
considered here, the system could be made symmetric by multiplying the
first equation in (\ref{eqn:Equations3x3}) by $-1$ and scaling
$\vec{P}$ and $\vec{B}$ by factors $c^2$ and $N^2$ respectively. Then
a MINRES method~\cite{Paige1975} with reduced memory requirements
could be used. However, this is not the case for general, spatially
varying values of the speed of sound and buoyancy frequency since
$\langle c^2 \phi,\nabla\cdot \delta\vec{u}\rangle\ne c^2 \langle
\phi,\nabla\cdot\delta\vec{u}\rangle$. To cover this more general
case, we therefore use GMRES to obtain our results.  Since the computationally expensive components in both Krylov methods are the operator application and preconditioner solve, the performance should be comparable and the only difference is the higher memory requirement of the GMRES method. For a discussion of the non-symmetry of the elliptic problems in atmospheric models we also refer the reader to~\cite{Thomas2003} where the authors argue that the GCR(k) algorithm (which is equivalent to GMRES) should be used.

To compare the two
preconditioners (Schur complement+geometric multigrid described in
section~\ref{sec:GeometricMGPreconditioner} vs.~PETSc
fieldsplit+BoomerAMG described in section~\ref{sec:PETScFieldsplit})
we first use the setup in Tab.~\ref{tab:GridSetup}. For comparison we
also use a single-level method (two iterations of Block-Jacobi
vertical line relaxation) since this is the approach used in many
atmospheric forecast models. For all numerical experiments we use five
levels in the geometric multigrid preconditioner. On each level one
pre- and one post-smoothing step with an overrelaxation parameter of
$\omega=0.8$ is employed. The well conditioned coarse grid problem (see
discussion at the end of section \ref{sec:tensor_product_multigrid}) is
solved approximately with two smoother iterations.

Instead of using a fixed number of coarse grid iteration, we could have reduced
the coarse grid residual to a fixed tolerance. However, then the
preconditioner is no longer guaranteed to be stationary and therefore a
flexible Krylov method must be used. Empirically, two iterations are good
enough.

We choose a typical Courant number of $\nuCFL=8.0$ for those runs (see section~\ref{sec:robustness} for the dependence on $\nuCFL$). Note, however, that the next-to-lowest order method resolves variations within one grid cell, so the effective Courant number is larger than $8.0$. After solving the system Eq.~\ref{eqn:PressureVelocitySystem} to the desired tolerance, buoyancy is reconstructed pointwise to obtain the solution of the full system Eq.~\ref{eqn:Equations3x3}. This requires an additional inversion of a well conditioned matrix.
\begin{table}
 \begin{center}
\begin{tabular}{lrrccc}
  \hline
  & \# horizontal & $\Delta x$ & \# vertical  & \multicolumn{2}{c}{number of unknowns} \\
& cells &                 & layers & per cell & total \\
  \hline\hline
    Lowest Order (LO) & 81,920 & $\approx90$km & 64 & 3.5  & 18.4 mio \\
    Next-to-Lowest Order (NLO) & 5,120  & $\approx360$km & 64 & 24 & 7.9 mio\\
    \hline
  \end{tabular}
  \caption{Grid setup for single-node tests. The total number of unknowns in the pressure and velocity space is shown in the two last columns.}
  \label{tab:GridSetup}
 \end{center}
\end{table}

All following results are obtained on the ARCHER supercomputer. Each
node consists of two 12 core Intel Xeon Ivybridge (E5-2697 v2)
processors, i.e.~24 cores in total. The spec sheet peak floating point
performance of a full node is $518.4\operatorname{GFLOPs/s}$ and the
maximum achieved memory bandwidth for the STREAM triad
benchmark~\cite{McCalpin1995} is $74.1\operatorname{GB/s}$. All code
was compiled using version 4.9.2 of the Gnu C compiler and suitable
flags are used to compile PETSc and generate optimised code.  All runs
were carried out on a full node (utilising all 24 cores) with MPI
parallelism.

Using threaded parallelism could potentially lead to further
performance enhancements by avoiding halo exchanges. Since the
relative cost of halo exchanges is suppressed by the surface-to-volume
ratio of the local domain and most of the work is spent on the finest
grid, it is unlikely that threaded parallelism will lead to
significant benefits for the grid sizes and low-order discretisations
considered in this work. In~\cite{Markall2013} an advection diffusion
equation is solved within the PyOP2 framework with different parallel
backends. The authors find that an OpenMP implementation is slightly
slower than the MPI equivalent.  This superiority of MPI over OpenMP
is also seen in other low-order studies, even when significant effort
is expended in tuning the OpenMP implementation, for
example~\cite{Reguly2016}.  While exploring shared memory and hybrid
parallelisation strategies would be interesting to improve the
performance of our solver in the strong scaling limit, it is beyond
the scope of this paper.

To obtain the following timings we do not include any overheads from
the just-in-time compilation of C-kernels in PyOP2. This is legitimate
since in a full model the linear equation will be solved a large
number of times and those overheads are completely amortised over the
model runtime.
\subsection{Algorithmic and computational performance}
\label{sec:alg-comp-perf}
\begin{figure}
 \begin{center}
   \includegraphics[width=0.9\linewidth]{\figdir/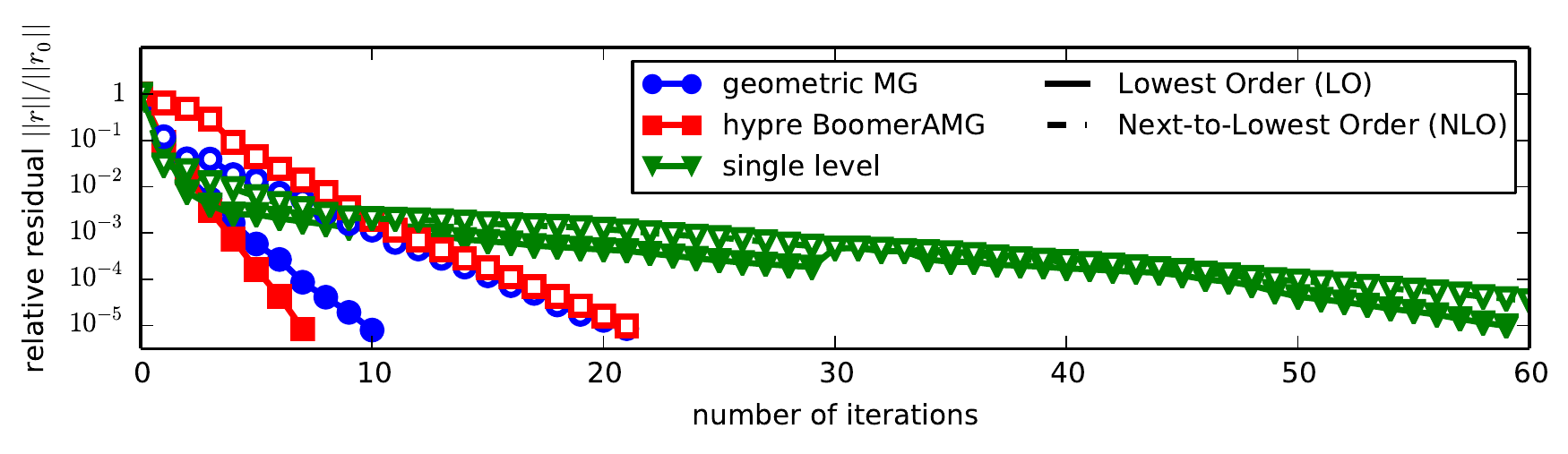}
   \caption{Convergence history for different solvers. Courant number is 8.0.}
   \label{fig:ConvergenceHistory}
 \end{center}
\end{figure}

Tab.~\ref{tab:SingleNodePerformance} shows a comparison of the
solution times, and the convergence history is plotted in
Fig.~\ref{fig:ConvergenceHistory}. At lowest order, the iterative
solver converges in ten or fewer iterations if a multigrid
preconditioner is used; the number of iterations is around 6 times
larger for the single level method. As can be seen from the
convergence history in Fig.~\ref{fig:ConvergenceHistory}, the
asymptotic convergence rate of the single level method (filled green
triangles) is significantly worse than for the multigrid methods
(filled blue circles and red squares).

\begin{table}
  \begin{center}
\begin{tabular}{llrrrr}
\hline
 & Preconditioner & $t_{\text{total}}$ & $t_{\text{setup}}$\footnotemark & $t_{\text{iter}}$ & $n_{\text{iter}}$\\
\hline\hline
\multirow{3}{*}{Lowest Order (LO)} &  geometric multigrid & 5.36 & 1.91 & 0.265 &  10\\
&  hypre BoomerAMG & 5.98 & 3.19 & 0.351 &   7\\
&  single level & 13.91 & 1.46 & 0.201 &  59\\
\hline
\multirow{3}{*}{Next-to-Lowest Order (NLO)} &  geometric multigrid & 11.07 & 3.85 & 0.307 &  21\\
&  hypre BoomerAMG & 12.17 & 4.55 & 0.353 &  21\\
&  single level & 22.71 & 3.52 & 0.230 &  81\\
\hline
\end{tabular}
 \end{center}
 \caption{Single node solver comparison for Courant number $\nuCFL=8.0$. The total solution time, setup time and time per iteration is shown together with the number of GMRES iterations which is required to reduce the residual by five orders of magnitude.}
 \label{tab:SingleNodePerformance}
\end{table}
\footnotetext{Due to technical reasons, the setup time for the AMG preconditioner includes the time for the first iteration.}
Although the time per iteration for the single level method is lower,
the additional iterations result in a significantly longer time to
solution. In total the single level method is more than twice as
expensive as a multigrid method, which is similar to the gain observed
in~\cite{Mueller2014a}. The fastest solution time is achieved with the
geometric multigrid preconditioner, which is about $12\%$ faster than
the AMG solver both at lowest order and at next-to-lowest order. While this
shows some gains from using a bespoke preconditioner, this is nowhere
near the $10\times$ speedup observed for the simplified setup
in~\cite{Mueller2014a}. Reasons for this are discussed in
section~\ref{sec:CommentsMatrixFree}. An initially surprising result
is the large setup time of the geometric multigrid preconditioner
which, although smaller than the AMG setup, is still significant. The
setup time for the single-level method is not much smaller than for
the multigrid algorithm, which implies that most of the time is spent
on the finest multigrid level. As detailed in
section~\ref{sec:breakdown} a significant proportion of this setup
time is taken up by assembly of the operators for the mixed system and
the Helmholtz operator on the finest multigrid level, which is
required for all preconditioners.

At next-to-lowest order, the asymptotic convergence rates of the multigrid
methods are again comparable, and both methods require 21 iterations
to converge. The asymptotic convergence rate of the single level
method is significantly worse and it needs almost 4 times as many
iterations as the geometric multigrid preconditioner. Again the total runtime
is twice as large for the single level method.
\subsection{Robustness}
\label{sec:robustness}
\begin{figure}[h]
 \begin{center}
  \begin{minipage}{0.45\linewidth}
    \includegraphics[width=\linewidth]{\figdir/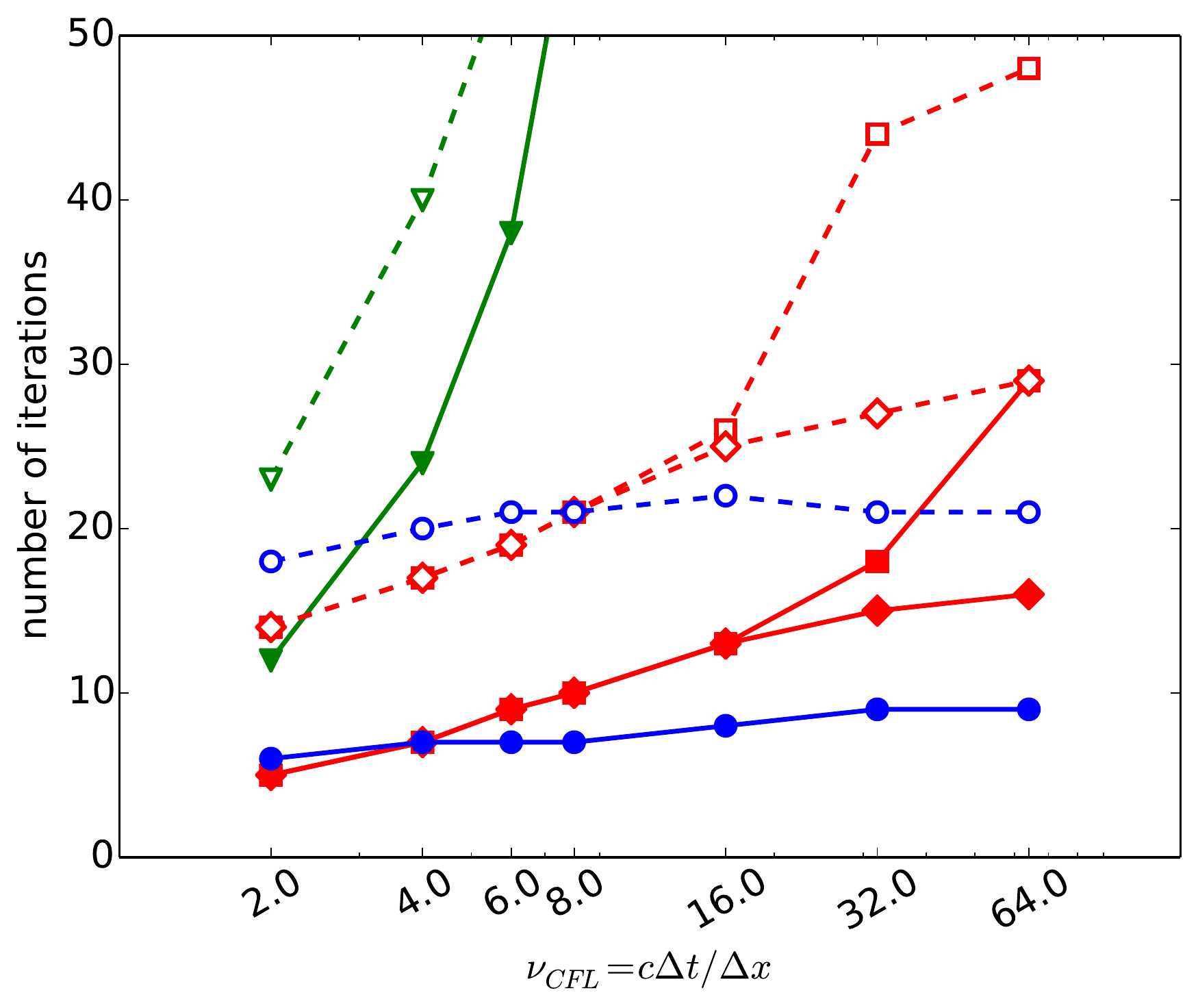}
  \end{minipage}
  \hfill
  \begin{minipage}{0.45\linewidth}
    \includegraphics[width=\linewidth]{\figdir/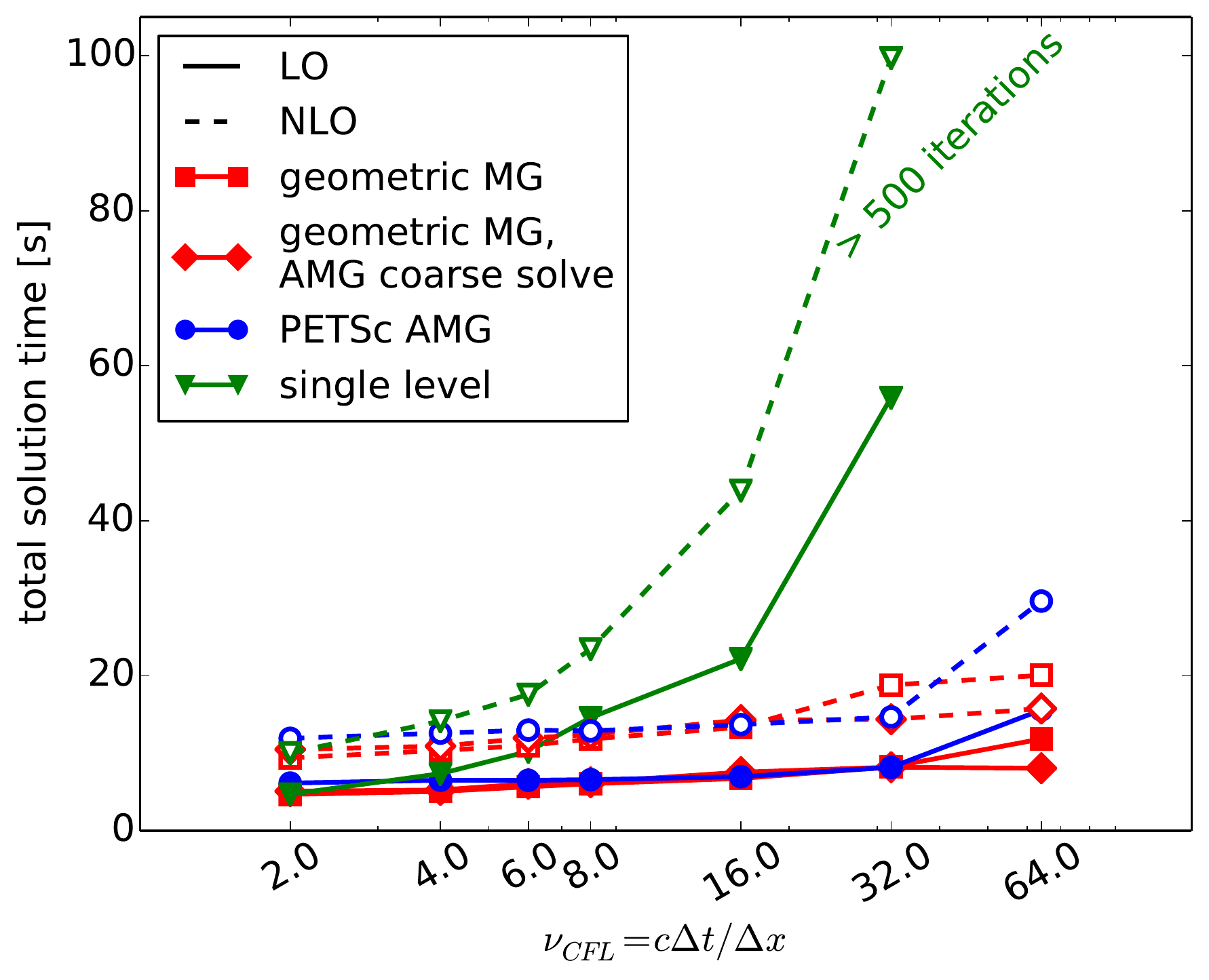}
  \end{minipage}
  \caption{Number of iterations (left) and total solution time (right) as a function of the Courant number. Results are shown both for Lowest Order (LO) and Next-to-Lowest Order (NLO). To solve the coarse grid problem in the geometric multigrid preconditioner we use either the default setup of two smoother iterations (squares) or one BoomerAMG V-cycle (diamonds).}
  \label{fig:CourantNumberVariation}
 \end{center}
\end{figure}

Of particular interest is the robustness of the solver under variations of the time step size and grid resolution. The number of iterations and total solution time as a function of the Courant number is shown in Fig.~\ref{fig:CourantNumberVariation}.
Note that the range we explore here is larger than what is typical in meteorological applications where $\nuCFL=\order(2-10)$. The number of iterations for the AMG preconditioner is virtually independent of the Courant number up to values of $\nuCFL=64$. The total solution time, however, increases sharply for the largest Courant number. The increase in the number of iterations for the geometric multigrid preconditioner can be partially explained by the fact that we use a fixed number of levels and the coarse grid problem becomes increasingly ill conditioned. To demonstrate this we include results in which the coarse grid problem of the geometric multigrid preconditioner is solved with a BoomerAMG V-cycle instead of two smoother iterations. In this configuration the increase in the number of GMRES iterations is very modest for $\nuCFL > 16$. As a result the total solution time is reduced for large Courant numbers. However, as expected, for small Courant numbers solving the coarse grid problem with a small number of smoother iterations works very well and it is not necessary to use a more powerful coarse grid solver. While the single-level method might be competitive for unrealistically small Courant numbers, the number of iterations increases dramatically with $\nuCFL$.

\begin{figure}
 \begin{center}
    \includegraphics[width=0.45\linewidth]{\figdir/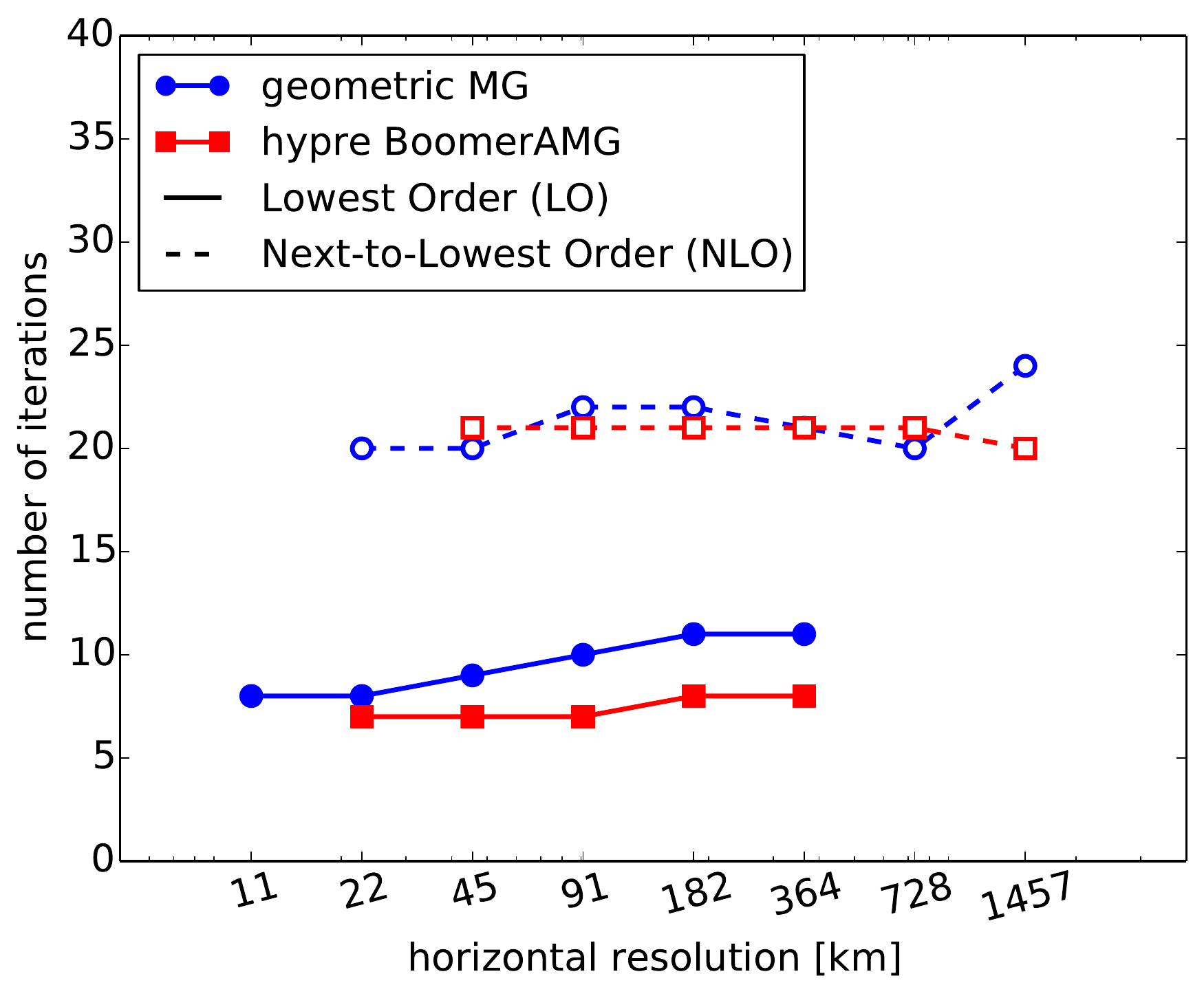}
    \hfill
    \includegraphics[width=0.5\linewidth]{\figdir/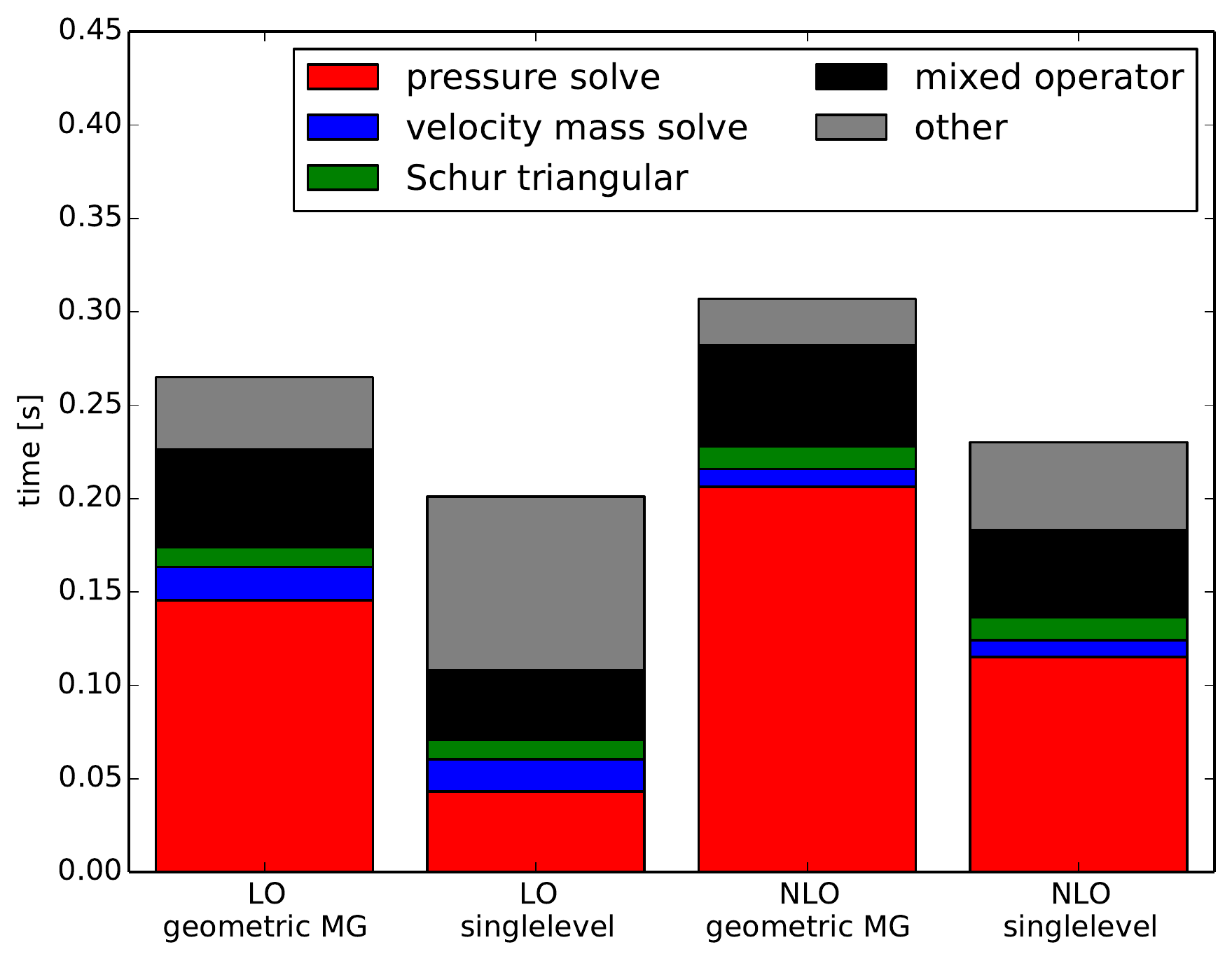}
  \caption{Number of iterations for increasing grid resolution (left) and breakdown of the time per iteration for different solvers (right).}
  \label{fig:AlgorithmicScaling}
 \end{center}
\end{figure}

In addition, it is important to check that the convergence rate of the multigrid solver is independent of the grid resolution.
To verify this, we varied the horizontal resolution and recorded the number of iterations, which is shown in Fig.~\ref{fig:AlgorithmicScaling} (left). For this test we kept the Courant number constant at $\nuCFL=8.0$, i.e.~as the resolution increases, the time step size decreases linearly with the horizontal grid spacing. We conclude that as expected the multigrid solvers show grid independent convergence.
\subsection{Breakdown of time per iteration}\label{sec:breakdown}
One possible concern is that the Firedrake-based implementation is significantly slower than a monolithic implementation in a fully compiled language, which would skew the comparison in runtimes reported in Tab.~\ref{tab:SingleNodePerformance}. To address this concern, we first remind the reader that the computationally most expensive loops over the grid are implemented by generating optimised C-kernels which are executed over the grid in PyOP2.  The performance of these finite element kernels in the Firedrake framework is studied in detail in~\cite{Luporini2015}; the performance of the Firedrake framework as a whole is discussed in~\cite{Rathgeber2015}, where it is shown that the overhead of using the high-level framework is negligible as long as there are more than $\mathcal{O}(5\times10^4)$ unknowns per MPI process (all results in this paper are obtained with $7.7\cdot 10^5$ (LO) or $3.3\cdot 10^5$ (NLO) unknowns per core).

To further quantify this in our case, we show a breakdown of the time per iteration for the multigrid- and single-level solvers in Fig.~\ref{fig:AlgorithmicScaling} (right) and report the times spent in the most important grid iterations in the last row of Tab.~\ref{tab:MultigridBreakdown}. As can be seen from the figure, in almost all cases (the exception being the lowest order single level method) more than half of the time is spent in the pressure solve, i.e. the multigrid or single-level preconditioner. Applying the mixed operator also takes up a significant amount of time, and the same operation is necessary for the PETSc fieldsplit+AMG implementation.

The time spent in the geometric multigrid preconditioner is further broken down by multigrid levels in Tab.~\ref{tab:MultigridBreakdown}. As expected from the ``total'' column, which gives the total time spent on a particular level, this is dominated by the finest level. Naively one would expect a reduction of cost by a factor of four from one level to the next coarser, since the horizontal grid is coarsened by a factor two in each direction. The only exception is the difference between level 4 and 3 in the next-to-lowest order case, which corresponds to a $p$-refinement step from $DG_1\otimes DG_1$ (6 unknowns per cell) to $DG_0\otimes DG_0$ (1 unknown per cell). The theoretical reduction factor is even larger than 6 since the local stencil size is reduced significantly when going from next-to-lowest order to lowest order on the same grid.
However the rates that are observed in practice are smaller.  This is due to a combination of worse communication to computation ratio on the coarser levels (they have a relatively larger halo) and the non-scalable fixed costs incurred by the Firedrake framework as detailed in~\cite{Rathgeber2015}.

On each level the most expensive components of the multigrid algorithm are the application of the operators $\hat{H}$ and $\hat{H}_z^{-1}$ in the residual calculation and the smoother application. Since we split the operator $\hat{H}$ into a horizontal and a vertical component, those operations can be expressed as
\begin{equation}
 \begin{aligned}
  \vec{v} &= \hat{H}\vec{u} = \hat{H}_z\vec{u}+\hat{H}_h\vec{u} &\text{(operator application)}\\
  \vec{u} &\mapsto \vec{u} + \hat{H}_z^{-1}(\vec{b}-\hat{H}\vec{u}) = 
\vec{u} + \hat{H}_z^{-1}(\vec{b}-\hat{H}_h\vec{u}-\hat{H}_z\vec{u})&\text{(smoother application)}
 \end{aligned}
\end{equation}
Since before the pre-smoother application the solution $\vec{u}$ is
zero, the total cost of those operations on on the finer multigrid
levels is
$(n_{\operatorname{pre}}+n_{\operatorname{post}})(\hat{H}_z+\hat{H}_h+\hat{H}_z^{-1})$. On
the coarsest level, where we apply the smoother
$n_{\operatorname{coarse}}$ times, the cost is
$(n_{\operatorname{coarse}}-1)(\hat{H}_z+\hat{H}_h)+n_{\operatorname{coarse}}\hat{H}_z^{-1}$. In
all numerical experiments we use
$n_{\operatorname{pre}}=n_{\operatorname{post}}=1$ and
$n_{\operatorname{coarse}}=2$.  The total cost of those operator
applications is listed in the last column. The remaining three columns
show the time for one individual application of $\hat{H}_h$,
$\hat{H}_z$ and $\hat{H}_z^{-1}$ on a particular multigrid level. This
confirms that indeed a significant amount of time is spent in the
PyOP2 parallel loops and we focus on analysing the performance of those parts
of the code. In the following we
construct a performance model for the operator application and confirm
that the implementation is indeed efficient: the achieved memory
bandwidth is close to the peak for the machine.
\begin{table}
  \renewcommand{\arraystretch}{1.25}
 \begin{center}
\begin{tabular}{|c|ccccc|ccccc|}
\hline
& \multicolumn{5}{|c|}{Lowest Order (LO)} & \multicolumn{5}{|c|}{Next-to-Lowest Order (NLO)}\\
level& total & $\hat{H}_h$ & $\hat{H}_z$ & $\hat{H}_z^{-1}$ & all $\hat{H}$
     & total & $\hat{H}_h$ & $\hat{H}_z$ & $\hat{H}_z^{-1}$ & all $\hat{H}$\\
\hline\hline
0  &   0.0076 &   0.0002 &   0.0008 &   0.0009 &   0.0027
   &   0.0072 &   0.0001 &   0.0007 &   0.0008 &   0.0025\\
1  &   0.0151 &   0.0003 &   0.0009 &   0.0010 &   0.0043
   &   0.0140 &   0.0002 &   0.0007 &   0.0009 &   0.0035\\
2  &   0.0174 &   0.0006 &   0.0012 &   0.0012 &   0.0060
   &   0.0150 &   0.0002 &   0.0008 &   0.0010 &   0.0040\\
3  &   0.0315 &   0.0022 &   0.0022 &   0.0020 &   0.0128
   &   0.0213 &   0.0023 &   0.0012 &   0.0013 &   0.0096\\
4  &   0.0683 &   0.0073 &   0.0053 &   0.0049 &   0.0350
   &   0.1461 &   0.0137 &   0.0080 &   0.0413 &   0.1261\\
\hline
total &   0.1399 &   0.0105 &   0.0105 &   0.0100 &   0.0609
      &   0.2035 &   0.0165 &   0.0115 &   0.0453 &   0.1457 \\
\hline
\end{tabular}
   \caption{Time spent on different levels of the multigrid V-cycle, as well as the key operations on each level. The last column labelled ``all $\hat{H}$'' shows the time spent in the PyOP2 loops for all operator applications.}
  \label{tab:MultigridBreakdown}
 \end{center}
\end{table}
\subsection{Performance model}\label{sec:PerformanceModel}
Iterative solvers employing assembled sparse operators are known to be
bandwidth-, rather than compute bound~\cite{Gropp2000}.  Assuming that
the vectors in a matrix vector product are cached perfectly,
multiplication by each non-zero entry streams one scalar and one
column index and performs one floating point multiply and one add.
This results of an arithmetic intensity of $2/(\operatorname{sizeof(\texttt{double})}+\operatorname{sizeof(\texttt{int})}) = 1/6$ flops per
byte.  Modern hardware delivers somewhere between 4 and 10 flops per
byte (recall that an ARCHER node provides 518.4 Gflop/s and a
bandwidth of 74.1 Gbytes/s, resulting in 7 flops/byte for a balanced
code).  With a good degree of freedom numbering, the assumption of
good vector caching is close to true, and high performance sparse
matrix implementations typically achieve an appreciable fraction of
STREAM bandwidth. It is therefore clear that if we want to improve
over this baseline, we can do little without changing the way we apply
matrix-vector products.

Rather than assembling the full operator, finite element solvers may
be implemented in a matrix-free manner by providing the matrix-vector
product directly.  This, along with the addition of a matrix-free
preconditioner is all iterative methods require.  Such approaches are
typically used for high-order discretisations (polynomial degree 4 and
upwards)~\cite{Brown2010}, although recent work shows that
significant speedup can already be achieved at degree two on
hexahedral meshes~\cite{May2014} when the tensor product structure of
the basis is exploited.  In contrast to these studies, we
are in a low-order regime (degree zero or one) and only have a partial
tensor product structure for the basis.  Additionally, since our
discretisation only couples degrees of freedom through faces, the
sparsity of the assembled operator is not too bad.

Finally, as discussed above, we apply the multigrid cycle to the Schur
complement operator $S$.  Since it contains a velocity mass matrix
inverse, its action is not purely expressible in terms of finite
element integrals.  Instead, we assemble (into a non-sparse banded
matrix) a column-wise decoupled approximation $\tilde{S}$ and apply
this action in the iterative solver.  For algorithmic
reasons we need to apply $\tilde{S}^{-1}$ exactly, which we do by
inverting the column-wise blocks.  Our method is therefore not truly
matrix-free, but instead exploits structure in the discretisation to
provide a more efficient storage format for the assembled operator: we
only need to stream matrix entries (rather than entries and column
indices) for a bandwidth saving of 33\% (50\% if using 64 bit integers
or single precision scalars).  To model performance, we count flops
and bytes for the application of the Schur complement.  The
block-diagonal mass-matrix is inverted once per solve, and we count
this as well.

\subsubsection{Schur complement matrix-vector products}
\label{sec:schur-complement-mat}
As quantified numerically in section~\ref{sec:breakdown}, the
computational bottleneck of the algorithm is application of the Schur
complement operator to a vector and the banded (or tridiagonal) matrix
solve in the vertical direction. We therefore concentrate our analysis
on these components.  The three key operations are the applications of
the operators $\hat{H}_h$, $\hat{H}_z$ and $\hat{H}_z^{-1}$. While
$\hat{H}_h$ is stored as a standard CSR sparse matrix, $\hat{H}_z$ is
stored as a banded matrix in the format described in
section~\ref{sec:BandedMatrixAlgebra}. At lowest order this matrix has
$m=64$ rows in each vertical column of the grid and is tridiagonal
with bandwidth $n_{\operatorname{BW}}=3$.  The next-to-lowest order matrix has
$m=384$ rows and bandwidth $n_{\operatorname{BW}}=27$. Since there are
6 degrees of freedom per cell, this bandwidth could be reduced to
$3\times 6=18$ if the banded matrix storage format were adapted to
account for the block-structure of the matrix. However, in the current
code this would prevent the use of LAPACK routines and should be
considered as an additional optimisation which will be explored in the
future.

To apply $\hat{H}_z$ to a vector, the input vector of size $m$ has to
be read from memory, an output vector of the same size has to be
written back and the banded matrix has to be loaded. The
total amount of data transferred is
\begin{equation}
  n_{\operatorname{bytes}}(\hat{H}_z) = m (n_{\operatorname{BW}}+2)\cdot\operatorname{sizeof(\texttt{double})}
= 8m (n_{\operatorname{BW}}+2).
\label{eqn:bytesHhatz}
\end{equation}
At lowest order we use a direct tridiagonal solver to apply
$\hat{H}_z^{-1}$, and the total amount of moved data is exactly the
same as in Eq.~\ref{eqn:bytesHhatz},
$n_{\operatorname{bytes}}(\hat{H}_z^{-1},\text{lowest
  order})=n_{\operatorname{bytes}}(\hat{H}_z)$. At next-to-lowest order we
calculate $\hat{H}_z^{-1}$ by an LU backsubstitution using the LAPACK
routine \verb+dgbtrs+. This requires additional storage of size $m$
for the pivot array and for additional $(n_{\operatorname{BW}}-1)/2$
matrix rows in the LU factorisation. Hence the total data volume is
\begin{equation}
\label{eqn:bytesHhatzinv_ho}
  n_{\operatorname{bytes}}(\hat{H}_z^{-1},\text{NLO}) = \frac{3}{2}m (n_{\operatorname{BW}}+1)\cdot\operatorname{sizeof(\texttt{double})}+m\cdot \operatorname{sizeof(\texttt{int})} = 4m(3n_{\operatorname{BW}}+4).
\end{equation}
Based on those estimates, we quantify the achieved memory bandwidth as follows: Since all data is contiguous in the vertical direction, it is reasonable to assume perfect caching, and we can estimate a ``useful'' memory bandwidth of
\begin{equation}
  \operatorname{BW}(A) = \frac{n_{\operatorname{bytes}}(A)\cdot n_{\operatorname{cell}}}{t(A)}\label{eqn:MemoryBW}
\end{equation}
where $t(A)$ is the time it takes to apply the operator $A$ and
$n_{\operatorname{cell}}$ is the number of vertical columns.  Note
that any less-than-perfect caching would manifest itself in a useful
bandwidth which smaller than the achievable peak memory bandwidth.

For the horizontal operator $\hat{H}_h$ let $M$ denote the total number of rows and $N_{\operatorname{nz}}$ the total number of nonzero entries. Following~\cite{Gropp2000} we need to read $M$ real numbers for the input vector and write back $M$ real numbers to store the output vector. Since the matrix is stored in compressed row storage, we also have to read $N_{\operatorname{nz}}$ real numbers as well as $M+N_{\operatorname{nz}}$ integers from memory.
The total amount of moved memory is in applying $\hat{H}_h$ to a
vector is therefore
\begin{equation}
\label{eqn:bytesHhath}
 \begin{aligned}
  n_{\operatorname{bytes}}(\hat{H}_h) &= (2M+N_{\operatorname{nz}})\cdot \operatorname{sizeof(\texttt{double})} + (M+N_{\operatorname{nz}})\cdot \operatorname{sizeof(\texttt{int})}
 = 20M+12N_{\operatorname{nz}}.
 \end{aligned}
\end{equation}
In analogy to Eq.~\ref{eqn:MemoryBW} the bandwidth can be calculated by dividing this number by the measured time.

Useful bandwidths calculated in this ways are reported in
Tab.~\ref{tab:usefulBandwidth}. On the finer levels, which amount for
most of the time, the achieved bandwidth corresponds to a significant
fraction of the achievable peak as measured using STREAM triad.
\begin{table}
\begin{center}
\begin{tabular}{|l|rrr|rrr|}
\hline
& \multicolumn{3}{|c|}{Lowest Order (LO)} & \multicolumn{3}{|c|}{Next-to-Lowest Order (NLO)}\\
multigrid level & $\hat{H}_h$ & $\hat{H}_z$ & $\hat{H}_z^{-1}$& $\hat{H}_h$ & $\hat{H}_z$ & $\hat{H}_z^{-1}$\\
\hline\hline
0 &   7.33 &   2.52 &   1.55 &   2.40 &   0.70 &   0.47\\
1 &  32.05 &   6.65 &   4.50 &   8.81 &   2.30 &   1.60\\
2 &  48.36 &  14.50 &   4.00 &  31.22 &   7.23 &   5.29\\
3 &  62.28 &  28.09 &  29.29 &  14.31 &   4.69 &   8.99\\
4 &  73.53 &  39.30 &  40.78 &  68.11 &  45.92 &  12.36\\
\hline
\end{tabular}
\caption{Memory bandwidth (in GB/s) for the core matrix operations in
  applying the preconditioner.  Data volume is calculated using the
  perfect cache models of
  Eqs.~\ref{eqn:bytesHhatz}--\ref{eqn:bytesHhath} as appropriate.  On
  the finest level of the grid we obtain upwards of 50\% of the STREAM
  triad bandwidth except for the application of $\hat{H}_z^{-1}$ at
  next-to-lowest order.}
\label{tab:usefulBandwidth}
\end{center}
\end{table}
\subsection{Breakdown of setup time}\label{sec:breakdownSetup}
Unlike a truly matrix-free geometric multigrid solver, a significant amount of time is spent in the setup of the geometric multigrid solver. The most significant fraction of time (1.094s at lowest order and 1.477s at next-to-lowest order) is spent in the assembly of the operator for the mixed system in Eq.~\ref{eqn:PressureVelocitySystem}; this assembly is also required for the PETSc fieldsplit+AMG preconditioner. While total time spent in banded matrix algebra (0.290ss at LO, 1.163s at NLO) is relatively small, the cost of assembling the horizontal derivative $D_h$ (0.254s at LO, 0.187s at NLO) and lumped mass matrices $\tilde{M}_u^h$ (0.085s at LO, 0.102s at NLO) and $\tilde{M}_u$ (0.204s at LO, 0.322s at NLO) are not negligible, as is the time for multiplication with $D_h$ (0.298s at LO, 0.551s at NLO), which requires irregular memory access on the horizontally unstructured grid.
\ifbool{PREPRINT}{ %
A more detailed breakdown of the setup time can be found in Tab.~\ref{tab:breakdownSetup} in appendix~\ref{sec:breakdownSetupTable}.
}{} 
\subsection{Comments on matrix-free implementations}\label{sec:CommentsMatrixFree}
The relatively modest speedup of the geometric multigrid solver compared to the AMG algorithm raises interesting questions.
The main reason for the large speedups of the geometric multigrid solver (compared to an AMG implementation) reported previously in~\cite{Mueller2014a} is the fact that the algorithm could be implemented in a matrix-free way for the simple finite-volume discretisation used there (we also used a matrix-free implementation for the slightly more complicated model equation in~\cite{Dedner2015}). This approach avoided storage of the matrix which was re-assembled on the fly. In contrast, for an AMG algorithm the matrix elements have to be loaded from memory which is expensive on modern computer architectures. As a consequence, 
in~\cite{Mueller2014a} geometric multigrid leads to a $10\times$ increase in performance compared to AMG. This, however, is only true if the matrix-free implementation for the application of linear operators is indeed faster than applying the assembled matrix. As we will now argue, this is not true anymore for the low-order finite element discretisations considered in this work.

In general, if an operator is to be applied $n$ times, the cost of the matrix-free (MF) and matrix-explicit (MX) implementation is
\begin{xalignat}{2}
  t^{(\operatorname{MF})}(n) &= n\cdot t_{\operatorname{apply}}^{(\operatorname{MF})}, &
  t^{(\operatorname{MX})}(n) &= t_{\operatorname{assemble}}+n\cdot t_{\operatorname{apply}}^{(\operatorname{MX})}
\end{xalignat}
where $t_{\operatorname{apply}}^{(\operatorname{MF})}$ and
$t_{\operatorname{apply}}^{(\operatorname{MX})}$ is the time for one
operator application and $t_{\operatorname{assemble}}$ is the time it
takes to assemble the operator into a sparse matrix. For the code in~\cite{Mueller2014a} we had $t_{\operatorname{apply}}^{(\operatorname{MF})}\ll t_{\operatorname{apply}}^{(\operatorname{MX})}$ and it is always cheaper to use the matrix-free implementation.

In contrast to this, we find that the matrix-free approach does not improve performance for the low order finite element discretisations used in this work, since our implementation has $t_{\operatorname{apply}}^{(\operatorname{MF})} \gg
t_{\operatorname{apply}}^{(\operatorname{MX})}$.

For the mixed operator in Eq.~\ref{eqn:PressureVelocitySystem} the corresponding timings are shown in Tab.~\ref{tab:matrixfreecomparison}. In both cases, one matrix-free operator application is about five times as expensive as an application of the corresponding assembled operator. 
The main reason for this is the much larger number of floating point operations. As can be seen from Tab.~\ref{tab:performance_mixed_operator}, one matrix-free operator application requires around $100\times$ more FLOPs than the corresponding matrix-explicit implementation. Although one ARCHER node can execute $R_{\text{peak}}/BW_{\text{peak}}=518.4/74.1 \approx 7$ floating point operations for each byte read from memory and the matrix-explicit implementation requires more memory traffic since the matrix has to be read in addition to the field vectors,
overall we expect the matrix-free code to be slower. More specifically, we can estimate the relative runtime of the (FLOP-bound) matrix-free and the (bandwidth bound) matrix-explicit code purely from counting memory references in Tab.~\ref{tab:performance_mixed_operator} as
\begin{equation}
  \rho = \frac{t_{\operatorname{apply}}^{(\operatorname{MF})}}{t_{\operatorname{apply}}^{(\operatorname{MX})}} \approx 
  \frac{(\text{\#FLOPs [MF]})/R_{\text{peak}}}{(\text{\# bytes moved [MX]})/BW_{\text{peak}}}.
\end{equation}
Since both implementations run at a sizeable fraction of the respective peak performance (see last two columns in Tab.~\ref{tab:performance_mixed_operator}), this estimate ($\rho=2.5$ for Lowest Order and $\rho=1.5$ at Next-to-Lowest Order) is in the same ballpark as the measured ratio of times.

At lowest order the assembly
cost is amortised after about 6 operator applications, at next-to-lowest order
it is amortised after 8 applications. The overhead from the matrix
assembly is further mitigated by the fact that parts of the assembled
mixed operator are reused at other places in the code. For example,
the lumped H(div) mass matrix can be extracted very cheaply once the
full mass matrix $M_2$ has been assembled.
\begin{table}
 \begin{center}
  \begin{tabular}{lrrr}
   \hline
   & $t_{\operatorname{assemble}}$ 
   & $t_{\operatorname{apply}}^{(\operatorname{MX})}$
   & $t_{\operatorname{apply}}^{(\operatorname{MF})}$\\
   \hline\hline
   Lowest Order (LO) & 1.094 & 0.053 & 0.235\\
   Next-to-Lowest Order (NLO) & 1.577 & 0.054 & 0.263\\
   \hline
  \end{tabular}
  \caption{Time spent in operator assembly and operator application
    for the mixed operator. Times for both the matrix-explicit (MX)
    and the matrix-free (MF) case are shown.}
  \label{tab:matrixfreecomparison}
 \end{center}
\end{table}

\begin{table}
 \begin{center}
\begin{tabular}{llrrrrr}
\hline
 & order & \#FLOPs & \#bytes & arithmetic & FP performance & bandwidth\\
 &       &         & moved & intensity & [GFLOPs/s] & [GByte/s]\\
\hline\hline
\multirow{2}{*}{matrix free} & LO & $4.8\cdot 10^{10}$ & $5.7\cdot 10^{8}$ & 84.38 & 203.3 & 2.2\\
 & NLO & $4.6\cdot 10^{10}$ & $2.0\cdot 10^{8}$ & 236.12 & 176.5 & 0.7\\
\hline
\multirow{2}{*}{matrix explicit} & LO & $3.7\cdot 10^{8}$ & $2.9\cdot 10^{9}$ & 0.13 & 7.0 & 51.8\\
 & NLO & $7.0\cdot 10^{8}$ & $4.5\cdot 10^{9}$ & 0.16 & 12.9 & 76.8\\
\hline
\end{tabular}
  \caption{Number of floating point operations and memory moved for the mixed operator.}
  \label{tab:performance_mixed_operator}
 \end{center}
\end{table}

Consequently in any tests that we carried out the matrix-explicit implementation of the geometric multigrid solver was much faster than a matrix-free version. As a result the performance of both the AMG and multigrid solver is limited by the speed with which the matrix and field vector can be loaded from memory, and with which the matrix can be assembled in the setup phase (explicit matrix assembly of $\hat{H}_z$ was also necessary for the LU decomposition that is required at next-to-lowest order). Since the matrix size is comparable in both cases, it is not surprising that the time per iteration and absolute solution time is comparable for both methods. This picture is likely to change at higher discretisation orders where sum factorisation techniques can improve performance dramatically: as shown in~\cite{Brown2010,Vos2010,Kronbichler2012,May2014} on modern chip architectures with a large FLOP-to-bandwidth ratio, sum-factorised matrix free implementations can be faster than memory bound implementations which assemble the operator and apply it in a sparse matrix representation. At high order, storing the assembled matrix can also require significantly more memory and limit the simulated problem size.
\subsection{Parallel scaling}
\begin{figure}[h]
 \begin{minipage}{0.45\linewidth}
  \includegraphics[width=1.0\linewidth]{\figdir/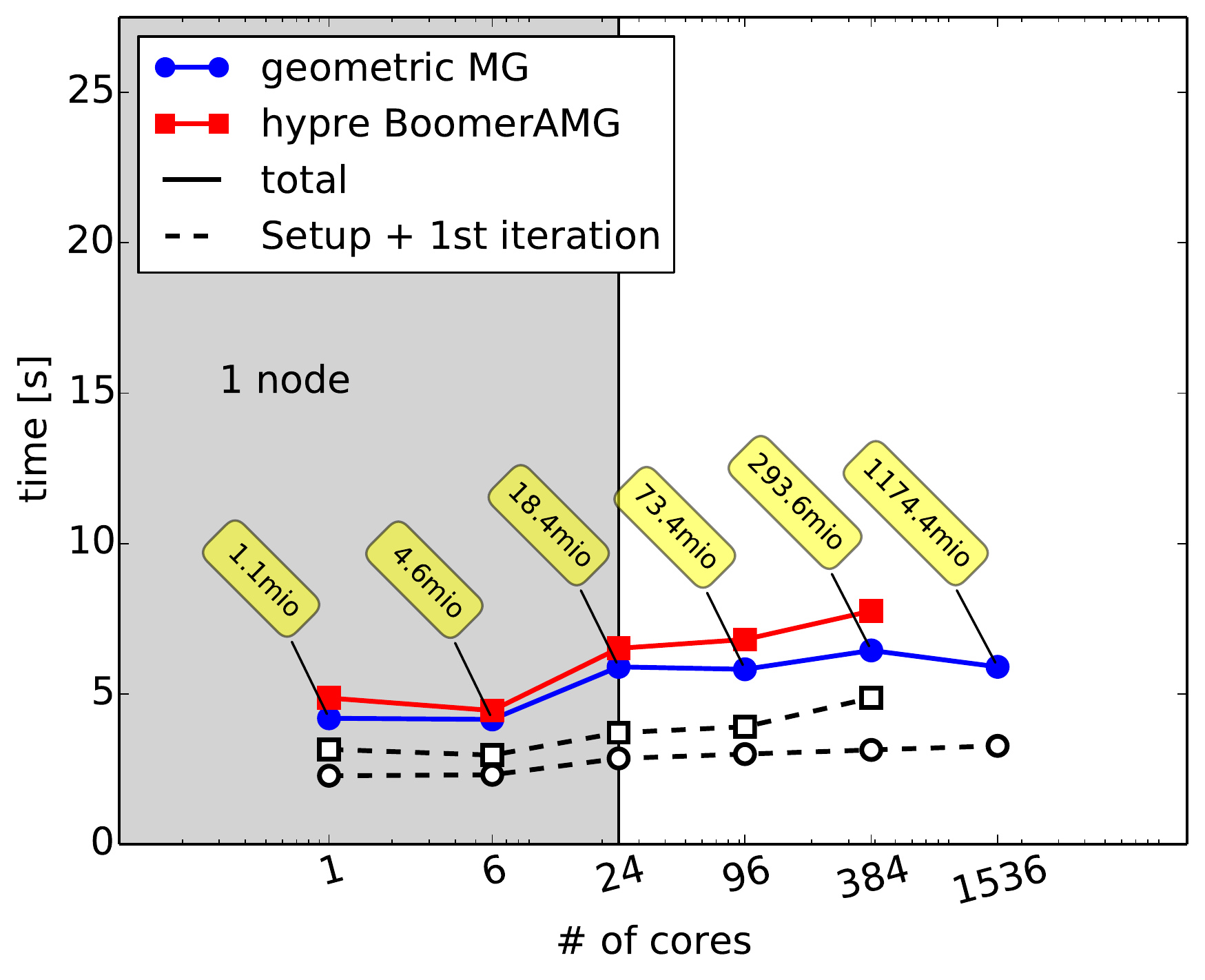}
 \end{minipage}
 \hfill
 \begin{minipage}{0.45\linewidth}
  \includegraphics[width=1.0\linewidth]{\figdir/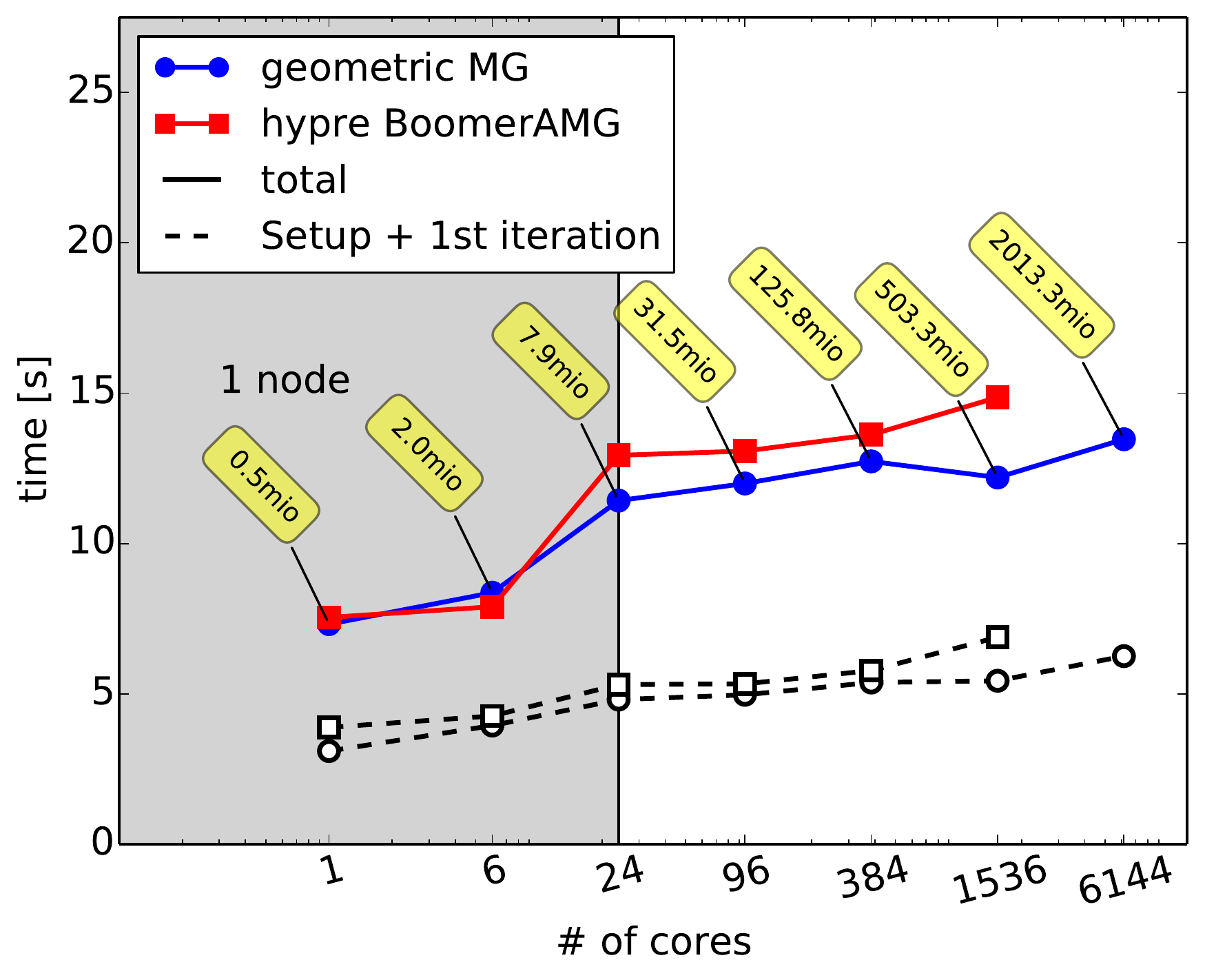}
 \end{minipage}
 \caption{Weak scaling of multigrid solvers on ARCHER. Results are shown for the lowest order (left) and next-to-lowest order (right) finite elements listed in Tab.~\ref{tab:FiniteElements}. Single-node data is shown with a gray background.}
 \label{fig:WeakScaling}
\end{figure}

The results of a weak scaling study of the multigrid solvers on ARCHER
is shown in Fig.~\ref{fig:WeakScaling}. As in the previous section we
fix the Courant number at $\nuCFL=8.0$, decreasing the timestep linearly
with higher grid resolution. The largest system solved at lowest order had $1.2\cdot 10^9$ unknowns on 1536 cores (64 full nodes); at next-to-lowest order we solved problems with up to $2.0\cdot 10^9$ degrees of freedom on 6144 cores (256 full nodes).
All solvers show excellent weak scaling once the nodes are fully populated. Any increases in runtime on partially occupied nodes can probably be attributed to the increasing saturation of the memory bandwidth as more cores are used.
\section{Conclusion}\label{sec:Conclusion}
The implementation of bespoke preconditioners for complex finite
element problems requires suitable abstractions which allow a
separation of concerns between the algorithm developer and the
computational scientist.  We discussed the implementation of
appropriate abstractions in the Firedrake finite element framework.
We then used the extended system to build a bespoke preconditioner for
the mixed finite element discretisation of a linear gravity wave
system in a thin spherical shell -- an important model system in
atmospheric flow applications.  Motivated by earlier work
in~\cite{Mueller2014a}, we constructed a bespoke tensor-product
geometric multigrid preconditioner, which is tailored to the strong
grid aligned anisotropy in the vertical direction. This preconditioner
results in a solver which is around $10\%$ faster than a purely algebraic
approach using PETSc fieldsplit preconditioning and hypre's algebraic
multigrid to solve the resulting elliptic problem.  For operationally
relevant CFL numbers, the multigrid preconditioners are about twice as
fast as the single level methods popular in current operational
models.  The different abstractions of the Firedrake/PyOP2 framework
simplify the implementation significantly: the user can express the
algorithm at the correct abstraction level, while still obtaining very
good performance. This is demonstrated by our careful analysis of key
components of the algorithm, which we show are running at a
significant fraction of the theoretical peak memory bandwidth. For
simpler finite volume or finite difference discretisations on
structured grids we previously demonstrated previously that it is
possible to achieve further speedups by using a matrix-free
implementation~\cite{Mueller2014a,Dedner2015}. However, here we show
that this is not possible for the low-order finite element
discretisations. In fact a matrix-based implementation based on a
BoomerAMG preconditioner achieves almost the same performance as our
bespoke geometric multigrid solvers.  The balance between matrix-free
and assembled operators may change with changes in future hardware,
and if we are able to exploit some of the structure in the
tensor-product basis (reducing the flop counts of the element
kernels); we intend to study this in the future.
\subsection{Future work}
There are several ways of extending the current work. In realistic atmospheric models orography will lead to a distortion of the grid and the decomposition of the velocity function space into a purely horizontal and purely vertical component is no longer valid, which can have an impact on the performance of the solver. Nevertheless the tensor-product multigrid algorithm, which assumes a perfect factorisation, can still be used as a good preconditioner, as has been confirmed by preliminary experiments (not reported here). Performance gains are also expected from a decomposition of the operators into tensor products. If, for example, an operator $A$ can be written as a tensor product $A=A_h\otimes A_z$, assembling (and applying) the operators $A_h$ and $A_z$ separately requires $\order(n)+\order(n_z)$ operations (and storage) instead of $\order(n\cdot n_z)$, and, since $n_z=\order(100)$ this will lead to significant speedups. This approach has been explored for a simplified problem in~\cite{Dedner2015}. In the finite element setting used in this work it would require a shallow-atmosphere approximation in the preconditioner.

Instead of using a Schur-complement factorisation and applying a multigrid preconditioner to the positive-definite system in (\ref{eqn:HelmholtzOperator}), one could also the multigrid algorithm for the full system in (\ref{eqn:PressureVelocitySystem}) or (\ref{eqn:Equations3x3}). This requires the construction of suitable smoothers, for example based on ``distributive iterations''~\cite{Wittum1989}. 

Ultimately the linear solver will be used inside a Newton iteration to
solve a non-linear problem in the full atmospheric model, and more
careful numerical studies will have to be carried out in this
context. While a non-linear multigrid (FAS) scheme~\cite[\S 8]{Brandt2011} could also be explored, the non-linearities in the problem considered here might not be large enough to justify this approach which requires computationally expensive non-linear smoothers. Guided by the work reported here, we are currently working on implementing both the PETSc fieldsplit preconditioner and the geometric multigrid preconditioner in the Fortran 2003 code base that is used to develop the Met Office's next generation dynamical core (codenamed ``GungHo'').
\section*{Acknowledgements}
This work was funded as part of the NERC project on Next Generation
Weather and Climate Prediction (NGWCP), grant numbers NE/K006789/1 and
NE/K006754/1.  LM additionally acknowledges funding from EPSRC grant
EP/M011054/1.  We gratefully acknowledge input from discussions with
our collaborators in the Met Office Dynamics Research group and the
GungHo! project.

This work used the ARCHER UK National Supercomputing Service
(http://www.archer.ac.uk).

All numerical experiments in this paper were performed with the
following versions of software, which we have archived on Zenodo:
Firedrake~\cite{zenodo_firedrake}; PyOP2~\cite{zenodo_pyop2};
FIAT~\cite{zenodo_fiat}; FFC~\cite{zenodo_ffc};
COFFEE~\cite{zenodo_coffee}; PETSc~\cite{zenodo_petsc};
petsc4py~\cite{zenodo_petsc4py}; UFL~\cite{zenodo_ufl}.  Additionally,
the simulation code itself is archived
as~\cite{zenodo_helmholtzsolver} and the performance data and plotting
scripts as~\cite{zenodo_data_plotting}.

\appendix
\ifbool{PREPRINT}{ 
\section{Derivation of model equations}\label{sec:ModelEquationDerivation}
Here we derive the equations Eq.~\ref{eqn:ContinuousEquations} from first principles. Large scale atmospheric flow is governed by the Navier Stokes equations and the equation of state:
\begin{xalignat}{4}
  \frac{D\vec{u}}{Dt} &= -c_p\theta \nabla \pi - g\hat{\vec{z}}, &
  \frac{D\theta}{Dt} &= 0, &
  \frac{\partial \rho}{\partial t}+\nabla\cdot\left(\rho\vec{u}\right) &= 0, &
  \rho\theta &= \Gamma \pi^\gamma.
\end{xalignat}
Expand around stationary profiles $\pi_0(z)$, $\theta_0(z)$ and $\rho_0(z)$
\begin{xalignat}{4}
  \vec{u}(\vec{x},t) &= \epsilon \vec{u}'(\vec{x},t), &
  \pi(\vec{x},t) &= \pi_0(z) + \epsilon \pi'(\vec{x},t), &
  \theta(\vec{x},t) &= \theta_0(z) + \epsilon \theta'(\vec{x},t), &
  \rho(\vec{x},t) &= \rho_0(z) + \epsilon \rho'(\vec{x},t)
\end{xalignat}
At lowest order in $\epsilon$ the stationary solution satisfies hydrostatic balance and the equation of state
\begin{xalignat}{2}
  c_p\theta_0 \partial_z \pi_0 &= -g, &
  \rho_0\theta_0 &= \Gamma\pi_0^\gamma.
\label{eqn:HydrostaticBalance}
\end{xalignat}
The equation of state also implies
\begin{xalignat}{2}
  \frac{\partial_z \rho_0}{\rho_0}
+ \frac{\partial_z \theta_0}{\theta_0}
&= \gamma\frac{\partial_z \pi_0}{\pi_0}, &
  \frac{\rho'}{\rho_0}
+ \frac{\theta'}{\theta_0} 
&= \gamma\frac{\pi'}{\pi_0}.\label{eqn:DifferentialEoS}
\end{xalignat} 
At $\order(\epsilon)$ the momentum equation and the conservation of potential temperature are
\begin{xalignat}{2}
  \frac{\partial\vec{u}'}{\partial t} &= c_p\theta_0 \nabla\pi'
  - c_p\hat{\vec{n}}(\partial_z\pi_0) \theta', &
  \frac{\partial \theta'}{\partial t} + \hat{\vec{n}}\cdot\vec{u}' (\partial_z \theta_0) &= 0.\label{eqn:epsilonMomentumBuoyancy}
\end{xalignat}
If we introduce the buoyancy $b$ and buoyancy (Brunt-Vais\"{a}l\"{a}) frequency $N$ with
\begin{xalignat}{2}
  b &\equiv g\frac{\theta'}{\theta_0},&
  N^2 &\equiv g\frac{\partial_z \theta_0}{\theta_0}
\end{xalignat}
these can be written (in the momentum equation we used Eq.~\ref{eqn:HydrostaticBalance} to replace $\partial_z\pi_0$ by $-g/(c_p\theta_0)$) as
\begin{xalignat}{2}
  \frac{\partial\vec{u}'}{\partial t} &= -c_p\theta_0 \nabla\pi' + \hat{\vec{n}} b\qquad \qquad\text{and}&
  \frac{\partial b}{\partial t} + N^2 \hat{\vec{n}}\cdot\vec{u}' &= 0.
  \label{eqn:BuoyancyEquation}
\end{xalignat}
The equation for the density is at $\order(\epsilon)$:
\begin{equation}
  \frac{\partial\rho'}{\partial t} + \hat{\vec{n}}\cdot\vec{u}'(\partial_z \rho_0) + \rho_0\nabla\cdot\vec{u}' = 0
\end{equation}
Divide by $\rho_0$ and use Eq.~\ref{eqn:DifferentialEoS} to replace $\rho'$ by $\pi'$ and $\theta'$
\begin{equation}
  \frac{\gamma}{\pi_0}\frac{\partial \pi'}{\partial t} 
+ \frac{\partial_z\rho_0}{\rho_0} \hat{\vec{n}}\cdot\vec{u}'-\frac{1}{\theta_0}\frac{\partial \theta'}{\partial t} + \nabla\cdot \vec{u}' = 0
\end{equation}
and then Eq.~\ref{eqn:epsilonMomentumBuoyancy} to eliminate the time derivative of $\theta'$
\begin{equation}
  \frac{\gamma}{\pi_0}\frac{\partial \pi'}{\partial t} 
+ \underbrace{\left(\frac{\partial_z\rho_0}{\rho_0}+\frac{\partial_z \theta_0}{\theta_0}\right) \hat{\vec{n}}\cdot\vec{u}'}_{A}+ \underbrace{\nabla\cdot \vec{u}'}_{B} = 0
\end{equation}
We now estimate the relative size of the two terms $A$ and $B$ in this equation. Define the scale height $D_f$ for a profile $f$ as
\begin{equation}
  D_f \equiv \min\left\{\left(\frac{\partial_z f_0}{f_0}\right)^{-1}\right\}.
\end{equation}
We now assume that the height $H$ of the domain is much smaller than all scale heights, $D_f\gg H$, i.e. the profiles vary only relatively slowly with height. Then we have 
\begin{equation}
  A \sim U\left(D_\rho^{-1}+D_\theta^{-1}\right) \ll UH^{-1} \sim B
\end{equation}
With this approximation the equation for the Exner pressure becomes
\begin{equation}
  \frac{\partial\pi'}{\partial t} + \frac{\pi_0}{\gamma}\nabla\cdot\vec{u}' = 0.
\end{equation}
Define
\begin{equation}
  p \equiv c_p \theta_0 \pi = c_p T_0 \frac{\pi'}{\pi_0}
\end{equation}
and note that the speed of sound $c$ is given by
\begin{equation}
  c^2 = \frac{c_p T_0}{\gamma}.
\end{equation}
When written in terms of the variable $p$ the pressure- equation and momentum equations become
\begin{xalignat}{2}
  \frac{\partial p}{\partial t} + c^2 \nabla\vec{u}' &= 0,
&
  \frac{\partial \vec{u}'}{\partial t} &= -\nabla p+\hat{\vec{n}}b.
  \label{eqn:PressureMomentumEquation}
\end{xalignat}
Here we have again used the condition on the scale heights to approximate
$\theta_0\nabla \pi' \approx \nabla(\theta_0\pi')$.
Together Eqs.~\ref{eqn:BuoyancyEquation} and \ref{eqn:PressureMomentumEquation} form the set of equations given in Eq.~\ref{eqn:ContinuousEquations}
}{} 
\ifbool{PREPRINT}{ 
\section{PETSc fieldsplit preconditioner options}\label{sec:fieldsplitoptions}
The options used for the fieldsplit preconditioner are shown in Tab.~\ref{tab:fieldsplitoptions}.
\begin{table}[h]
 \begin{center}
  \begin{tabular}{ll}
   \hline
   option & value\\
   \hline\hline
     \texttt{pc\_type} & \texttt{fieldsplit} \\
     \texttt{pc\_fieldsplit\_type} & \texttt{schur} \\
     \texttt{pc\_fieldsplit\_schur\_fact\_type} & \texttt{FULL} \\
     \texttt{pc\_fieldsplit\_schur\_precondition} & \texttt{selfp} \\
     \texttt{fieldsplit\_0\_ksp\_type} & \texttt{preonly} \\
     \texttt{fieldsplit\_0\_pc\_type} & \texttt{bjacobi} \\
     \texttt{fieldsplit\_0\_sub\_pc\_type} & \texttt{ilu} \\
     \texttt{fieldsplit\_1\_ksp\_type} & \texttt{preonly} \\
     \texttt{fieldsplit\_1\_pc\_type} & \texttt{hypre} \\
     \texttt{fieldsplit\_1\_pc\_hypre\_type} & \texttt{boomeramg} \\
     \texttt{fieldsplit\_1\_pc\_hypre\_boomeramg\_max\_iter} & 1 \\
     \texttt{fieldsplit\_1\_pc\_hypre\_boomeramg\_agg\_nl} & 0 \\
     \texttt{fieldsplit\_1\_pc\_hypre\_boomeramg\_coarsen\_type} & \texttt{Falgout} \\
     \texttt{fieldsplit\_1\_pc\_hypre\_boomeramg\_smooth\_type} & \texttt{Euclid} \\
     \texttt{fieldsplit\_1\_pc\_hypre\_boomeramg\_eu\_bj} & 1 \\
     \texttt{fieldsplit\_1\_pc\_hypre\_boomeramg\_interptype} & \texttt{classical} \\
     \texttt{fieldsplit\_1\_pc\_hypre\_boomeramg\_P\_max} & 0 \\
     \texttt{fieldsplit\_1\_pc\_hypre\_boomeramg\_agg\_nl} & 0 \\
     \texttt{fieldsplit\_1\_pc\_hypre\_boomeramg\_strong\_threshold} & 0.25 \\
     \texttt{fieldsplit\_1\_pc\_hypre\_boomeramg\_max\_levels} & 25 \\
     \texttt{fieldsplit\_1\_pc\_hypre\_boomeramg\_no\_CF} & \texttt{False}\\
   \hline
  \end{tabular}
 \caption{Options for PETSc fieldsplit preconditioner}
 \label{tab:fieldsplitoptions}
 \end{center}
\end{table}
\section{Breakdown of setup time}\label{sec:breakdownSetupTable}
A detailed breakdown of the setup time is shown in Tab.~\ref{tab:breakdownSetup}.
\begin{table}[h]
  \begin{center}
\begin{tabular}{lrr}
\hline
 & \multicolumn{2}{c}{order}\\
 & LO & NLO\\
\hline
\hline
assemble $D_h^T$ &  0.254 &  0.187\\
assemble $\tilde{M}^h_2$ &  0.085 &  0.102\\
assemble $\tilde{M}_2$ &  0.204 &  0.322\\
assemble operator for mixed system &  1.094 &  1.577\\
bandedmatrix: LU factorise ($A=LU$) & ---  &  0.089\\
bandedmatrix: add ($C=A+B$) &  0.029 &  0.176\\
bandedmatrix: columnwise assembly &  0.035 &  0.091\\
bandedmatrix: generate UFL form &  0.098 &  0.104\\
bandedmatrix: local assembly &  0.076 &  0.079\\
bandedmatrix: multiply ($C=A\cdot B$) &  0.019 &  0.043\\
bandedmatrix: multiply transpose ($C=A^T\cdot B$) &  0.033 &  0.581\\
multiply $D_h \times \tilde{M}_{2,\operatorname{inv}}^h D_h^T$ &  0.298 &  0.551\\
\hline
total  &  1.912 &  3.851\\
\hline
\end{tabular}
\caption{Breakdown of setup time. All times are given in seconds. Since a tridiagonal solver is used at lowest order, no time is spent in the LU factorisation.}\label{tab:setuptime}
\label{tab:breakdownSetup}
  \end{center}
\end{table}
\clearpage
}{} 
\bibliographystyle{elsarticle-num}

\end{document}